\documentclass[journal]{IEEEtran}
\ifCLASSINFOpdf
   \usepackage[pdftex]{graphicx}
   \graphicspath{{../pdf/}{../jpeg/}}
   \DeclareGraphicsExtensions{.pdf,.jpeg,.png}
\else
   \usepackage[dvips]{graphicx}
   \graphicspath{{../eps/}}
   \DeclareGraphicsExtensions{.eps}
\fi
\ifCLASSOPTIONcompsoc
  \usepackage[caption=false,font=normalsize,labelfont=sf,textfont=sf]{subfig}
\else
  \usepackage[caption=false,font=footnotesize]{subfig}
\fi

\usepackage{setspace}
\begin{document}
%
\title{Recent Advances and Challenges in Ubiquitous Sensing}
%
%
%

\author{Stephan~Sigg,~\IEEEmembership{Member,~IEEE,}
        Kai~Kunze,~\IEEEmembership{Member,~IEEE,}
        and~Xiaoming~Fu,~\IEEEmembership{Member,~IEEE}
\thanks{S. Sigg and X. Fu are with the Department of Computer Science and Mathematics, Georg-August University Goettingen, Goettingen, Germany,
e-mail: \{stephan.sigg,fu\}@cs.uni-goettingen.de.}
\thanks{K. Kunze is with the Department of Computer Science and Intelligent Systems of the Osaka Prefecture University, Osaka, Japan.}}

%
%

\markboth{ }%
{Sigg \MakeLowercase{\textit{et al.}}: Advances in Activity recognition}
%



\maketitle

\begin{abstract}
Ubiquitous sensing is tightly coupled with activity recognition.
This survey reviews recent advances in Ubiquitous sensing and looks ahead on promising future directions. 
In particular, Ubiquitous sensing crosses new barriers giving us new ways to interact with the environment or to inspect our psyche.
Through sensing paradigms that parasitically utilise stimuli from the noise of environmental, third-party pre-installed systems, sensing leaves the boundaries of the personal domain.
Compared to previous environmental sensing approaches, these new systems mitigate high installation and placement cost by providing a robustness towards process noise.
On the other hand, sensing focuses inward and attempts to capture mental activities such as cognitive load, fatigue or emotion through advances in, for instance, eye-gaze sensing systems or interpretation of body gesture or pose.
This survey summarises these developments and discusses current research questions and promising future directions.
\end{abstract}

\begin{IEEEkeywords}
Ubiquitous sensing, Activity recognition, Device-free, sentiment sensing, Pervasive Computing, RF signals, 
\end{IEEEkeywords}

%
\IEEEpeerreviewmaketitle

\section{Introduction}
With the stark penetration by smart and mobile devices, we continuously carry sensors of all kinds with us, which monitor every location, situation and activity.
More and more applications are exploiting these capabilities. 
Google Now, fourSquare, Facebook, Twitter and others gather, analyse and exploit large amounts of instantaneous, personalised information.
With this data, we can provide novel, intelligent and personalised services to the users.

\begin{figure}
\subfloat[Device- and individual-focused sensing of directly observable states]{\includegraphics[width=\columnwidth]{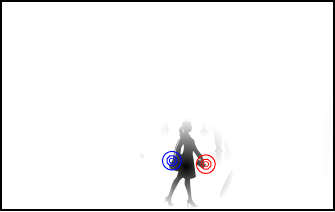}\label{figureTeaserA}}

\subfloat[Parasitic- and Sentiment sensing]{\includegraphics[width=\columnwidth]{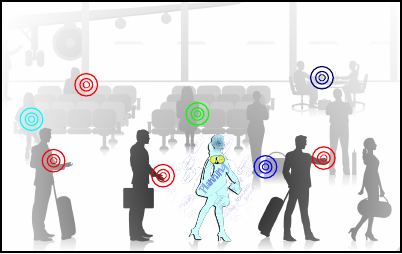}\label{figureTeaserB}}
\caption{Classical and future Ubiquitous sensing paradigms}
\label{figureTeaser}
\end{figure}

Development divisions in industry are currently exploring these possibilities, while research is evolving towards new frontiers; we see two main directions of this development:
\begin{description}
 \item[\textbf{\underline{Parasitic sensing}}]:\\ The parasitic utilisation of environmental, ubiquitously available sources in contrast to sensors on isolated, personal devices.
 \item[\textbf{\underline{Sentiment sensing}}]:\\ Interpreting sensor information to recognize mental states, intention, attention emotion and cognitive activities of individuals.
\end{description}
As depicted in figure~\ref{figureTeaser}, in traditional Ubiquitous Sensing, the focus of the sensing system lies on the status of a mobile, personal device or sensors attached to an individual and on this individual's directly observable actions (figure~\ref{figureTeaserA}).
The environment (surroundings, crowd, situations) are typically not covered by personal device sensors.
Consequently, the device is in a sense short-sighted with its perception limited to an isolated individual. 
However, considering a complete individual with her plans, emotions, intentions and mental states, classical sensing captures only the surface of that complex human system. 
Gradually, this focus is shifting towards the recognition of mental states, intention or emotion of individuals while increasingly environmental sensing sources are employed which combine zero installation cost with ubiquitous availability.
Only recently, a special issue of the IEEE Pervasive Computing magazine focused on the recognition of attention via sensing modalities~\cite{Pervasive_Ferscha_2014}.

As indicated in figure~\ref{figureTeaserB}, the new sensing paradigms extend the sensing range twofold: on the one hand, through the utilisation of environmental sources, additional and more fine-grained information on situations and surrounding entities is available. 
On the other hand, additional information on mental states can be derived. 

Parasitic sensing utilises environmental, ubiquitous sensing sources such as, for instance, audio or radio frequency~\cite{RFsensing_Pu_2013,Pervasive_Adib_2013,Cryptography_Schuerman_2011,Cryptography_Madiseh_2008} and thereby extends the perception of the sensing system beyond the boundaries of an individual device or person. 
Through the utilisation of stimuli from already installed infrastructure, coverage is maximised while installation cost is minimised.

Sentiment sensing focuses on people's mental state, intention or emotion, for instance, by interpreting eye-gaze information~\cite{Pervasive_Ishimaru_2014,Pervasive_Kunze_2013}, body gesture or pose~\cite{Pervasive_Castellano_2008,Pervasive_Jaggarwal_2012} and thereby directs and extends the perception of a sensing system inwards.

In this survey, we detail current advances towards parasitic and sentiment sensing and discuss open research challenges and promising future directions.

\section{Overview}
This section briefly sketches recent development that will foster and induce Parasitic and Sentiment Sensing.
Then, in section~\ref{sectionDeviceFree} and section~\ref{sectionCognitive}, current advances in these directions are detailed before, in section~\ref{sectionFuture}, lively discussed topics and future directions are introduced.

\subsection{The route to Parasitic Sensing}
Over the last decade, we have seen remarkable progress in the recognition of human activities or situations~\cite{Pervasive_Roggen_2009,Pervasive_Cheng_2013,Pervasive_Chen_2012,Pervasive_Chen_2012-2}.
This was driven by several strong developments in related areas.
First of all, sensing hardware has been greatly improved (e.g. size, accuracy and also new sensing modalities and sense-able quantities), enabling an enhanced perception of the world through sensors.
Also, machine learning has celebrated great successes (algorithms, toolboxes) and has become a mainstream ability that attracts a huge user base towards activity recognition.
Furthermore, rapid development in wireless protocols and near-global coverage of some technologies (e.g. UMTS, LTE) enabled the transmission at higher data rates and new usage areas through wireless communication.
Last, but not least, novel applications have spread that promote the publishing and sharing of all kinds of data (e.g. Facebook, Line, WhatsApp), which led to novel valuable inputs for activity recognition.

Even given this progress and innovation already, the field is on the verge towards a disruptive next change that will revolutionise usage patterns and open a multitude of new research directions.

Activity recognition in Ubicomp is going towards Big Data with systems developing capabilities to monitor virtually everybody, everywhere and without specifically installing system components at any particular physical location.

\begin{figure}
 \centering
 \includegraphics[width=\columnwidth]{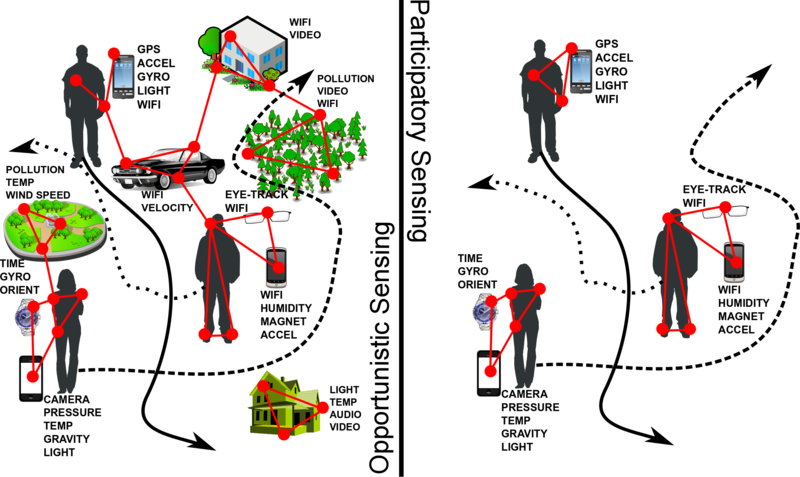}
 \caption{Participatory and Opportunistic sensing paradigms}
 \label{figureParticipatoryOpportunisticSensing}
\end{figure}

Fostered through the advancing Internet of Things and fueled by Opportunistic and Participatory Sensing campaigns (cf. figure~\ref{figureParticipatoryOpportunisticSensing}), we have been able to follow this development in recent years.

Opportunistic sensing has been viewed as one likely future of sensing~\cite{Opportunistic_Campbell_2008}.
Distributed devices provide their sensing capabilities to neighbouring devices, that are then empowered to access the remotely sensed information or to generate tasks for remote devices to acquire and share this information~\cite{OpportunisticSensing_Lane_2008,OpportunisticSensing_Kapadia_2008}.
This is a promising concept which greatly extends the perception of a mobile device to the joint perception of its neighbouring devices and environment.
In the frame of the OPPORTUNITY project~\footnote{Opportunity Project website: http://www.opportunity-project.eu/ (Mai 2014)}, an architecture for opportunistic sensing, in particular activity recognition was developed~\cite{OpportunisticSensing_Roggen_2009,OpportunisticSensing_Kurz_2011}.
However, we did not see a broad application and utilisation of Opportunistic sensing yet.

Opportunistic Sensing rises a number of issues not only regarding the mere technical implementation, protocols, mobility and timing.
It also touches aspects of privacy and security when alien devices are allowed to access potentially privacy-related personalised information in an uncontrolled manner~\cite{OpportunisticSensing_Kapadia_2008,OpportunisticSensing_Shin_2011}.
In particular, the concept envisions that arbitrary sensors can be accessed so that, apart from the also tremendous challenge to enable the seamless interaction technically, the design of a privacy or security preserving scheme is a nightmare which, with the sheer infinite possibilities and security threats posed by all the sensors, can hardly be solved.

With the proposal of Participatory Sensing~\cite{ParticipatorySensing_Burke_2006}, the privacy issues of Opportunistic sensing are solved pragmatically.
In this sensing principle, remote sensing is restricted to user-controlled mobile devices.
Remote devices are still expected to task neighbouring devices for sensed information, but human interaction is required in order to approve such request~\cite{OpportunisticSensing_Lane_2008}.
Consequently, not only is the range of devices restricted to explicitly user-controlled devices with an interactive interface but also the important principle of calmness and unobtrusiveness in Pervasive Computing is disregarded.
Instead, the mental load for a user with a Participatory Sensing Device is likely significantly increased as she might be frequently interrupted for interaction.

However, these developments indicate the direction in which activity recognition and sensing as a whole develop.
Instead of utilising device-bound sensors with limited range, future sensing will incorporate increasingly environmental sensing sources which have the potential to extend the perception of a sensing device beyond its physical boundaries.
Consequently, as discussed above, the reliance on explicit hardware sensors in the environment introduces communication overhead as well as technical, privacy and security-related problems.
As long as there is no real incentive for device-owners to make sensors on their devices available to the public, they will rather choose to protect their security and privacy as well as their battery by granting exclusively local access to sensors on a device.

A less problematic and yet simpler way to extend the perception of a mobile device into the environment is the utilisation of environmental stimuli that can be extracted from the noise of other systems.
The parasitic usage of environmental noise has been demonstrated by infrastructure mediated sensing paradigms~\cite{Pervasive_patel_2008,Pervasive_patel_2008b}, audio-based~\cite{ContextAwareness_Kunze_2007} and radio-frequency based approaches~\cite{RFSensing_Sigg_2014,RFsensing_Pu_2013} as detailed in section~\ref{sectionDeviceFree}.
We believe that the greatest potential underlies the RF-based systems since (A) RF is available ubiquitously (free frequency spectrum is sparse all over the world), (B) virtually all contemporary electronic devices incorporate an interface to the radio channel and (C) novel technical developments such as OFDM (cf. section~\ref{sectionFuture}) incorporate properties that will likely lead to better recognition accuracies on cheap off-the-shelf consumer devices.

This development is already under way with the community increasingly considering device-free techniques that relieve the monitored individuals from the burden of actually wearing any sensing hardware; and this evolution will continue in the direction of passive, device-free systems which exploit parasitic operation by re-using noisy emissions from ubiquitously available, environmental third-party pre-installed technology.

\subsection{The Route to Sentiment Sensing}
Activity Recognition started out with detecting very simple
physical states, walking, sitting, standing -- modes of locomotion -- in the 1990s. We came a long way from these simple classes to tracking a lot of high level activities, like car repair,
furniture assembly and Kung Fu exercises~\cite{Antifakos:2002p8030,heinz2006uws}.

The dedicated sensor systems used in the labs were not easily deployable. 
Yet, this changed with the advent of the smart phone as general computing platform. Suddenly "cheap" motion sensors
were available to everybody. Still using smart phones or
other consumer devices brought also new challenges.
The position and orientation of the devices was no longer fixed.
One had to cope with location and orientation changes
of the sensors~\cite{kunze2008dsd}.

Next we saw a push towards physiological sensing, first in the
medical application domain then also for more and more
sports and fitness research.

Now, more and more people get interested in the brain and brain
functions. We gather rich information in cognitive science,
psychology medicine and related fields about cognitive
processes. Therefore, we have now a sufficient basis to explore
cognitive task tracking in everyday life~\cite{kunze2013activity}.

The first impacts are already visible in the medical domain.
Here we see that sensor data from smart phones can predict depression episodes in patients with mental illnesses.
Motion data seems to correlate well with some mental states.
The same holds for the physiological data. Heart rate, blood oxygen level etc. can tell a lot about our cognitive condition especially combined with motion sensors
(e.g. if a user doesn't move much and his heart rate is increased, it could signify that he's excited)~\cite{matthews2007wearable}.

Yet, more interestingly, there are a couple of sensor modalities to track brain activity (in)-directly and we see them more and more embedded in consumer devices (e.g. the emotive headset to track brain activity using EEG).

It seems obvious to track brain activity directly using EEG or other brain imaging technologies. However, these technologies have severe limitations; either they are quite expensive and bulky (e.g. magnetic resonance imaging) or they require heavy filtering and analysing. 
As our skull is quite thick, brain signals are easily overshadowed by motion artifacts etc. 

One promising alternative is to use eye tracking, as gaze is directly correlated to some of the higher brain functions. 
There are two common approaches.
Optical eye tracking uses infra-red lights and camera to track the pupil.
Electrooculography uses electrodes to track eye movements, as our eye is a dipole~\cite{Pervasive_bulling_2009}.
\section{Device Free/ Radio Sensing} \label{sectionDeviceFree}
Sensing modalities for activity recognition or monitoring differ in their installation effort and range (cf.~figure~\ref{figureSensingModalities}).
\begin{figure*}
\textbf{\underline{Device-bound}}\\[.2cm]
\fbox{
  \begin{minipage}{.975\textwidth}\begin{scriptsize}\begin{sf}\begin{spacing}{0.9}
    \textit{\underline{Inertial sensors}}\\[.1cm]
    Accelerometer devices are becoming rapidly ubiquitous in modern day technology~\cite{Pervasive_Chang_2009,Pervasive_Laerhoven_2004}. 
    Employed for a broad range of use cases from mere environmental adjustment of devices to the recognition of individual user's situation\cite{Pervasive_Miluzzo_2008,Pervasive_Laerhoven_2008,Pervasive_Welbourne_2005}.
    Multiple sensors instrumented at multiple body locations utilised to recognise different activities~\cite{5952,Pervasive_Lester_2006,Pervasive_Tapia_2007,Pervasive_Laerhoven_2004}.
    Other related sensors are vibration sensors~\cite{Pervasive_Gordon_2010b,Pervasive_Laerhoven_2004}, or magnetic resonant coupling~\cite{Pervasive_Pirkl_2008}. 
    \\[.2cm]
    \textit{\underline{Bio-sensors}}\\[.1cm]
      Sensors to monitor the heart rate are employed to predict physical activity~\cite{Pervasive_Yannakakis_2008,Pervasive_Intille_2003,Pervasive_Michahelles_2003}.
      Popular in health related applications is also the monitoring of blood pressure~\cite{Pervasive_Christopoulou_2005} or electrocardiography ECG~\cite{Pervasive_seo_2004,Pervasive_lo_2005}. In addition, Electromygraphy(EMG) sensors are used to monitor the health status~\cite{Pervasive_gamecho_2013} or, e.g. facial EMG to support eye-gaze tracking sensors~\cite{Pervasive_bulling_2009}.
      This sensor class is feasible to record muscle activity (surface EMG electrodes)~\cite{Pervasive_Amft_2009,Pervasive_Ogris_2007}\\[.2cm]
         \textit{\underline{RF-based}}\\[.1cm]
      Device localisation is possible by employing WiFi signal strength and signal-to-noise ratio~\cite{Pervasive_Bahl_2000}, signal strength information from the active set at a GSM terminal~\cite{Pervasive_Otsason_2005,Pervasive_Varshavsky_2007}, or also via signal strength information of a set of signals received from nearby FM radio stations~\cite{Pervasive_Krumm_2003,Pervasive_Youssef_2005}.
      \end{spacing}
     \end{sf}
	\end{scriptsize}
  \end{minipage}
  }\\[.2cm]

\begin{minipage}{\textwidth}
  \textbf{\underline{Device-free}\\[.2cm]}
  \fbox{
    \begin{minipage}{.303\textwidth}\begin{scriptsize}\begin{sf}\begin{spacing}{0.9}
      \textbf{Installation-based}\\[.2cm]
      \textit{\underline{Video}}\\[.1cm]
      \linespread{0.5}Recognition of activities from video can reach remarkable accuracies~\cite{Pervasive_Chaquet_2013}. 
      Activities are identified via matching of templates, neighbour based or via statistica modelling~\cite{ActivityRecognition_Aggarwal_2011,LocationTracking_Cai_1998}. 
      However, video has high installation cost, is strictly range limited, fails in darkness and may violate privacy.\\[.2cm]
      \textit{\underline{Infrared}}\\[.1cm]
      Capturing of radiated infrared waves emitted from objects. 
      Infrared can be employed as imaging technology similar to video but with the benefit that human motion can be easily detected from the background regardless of the lighting conditions and colors of the human clothing and surfaces~\cite{Pervasive_Han_2005,5836}.
      The technique is limited in sensing range and requires careful and dense deployment. \\[.2cm]
      \textit{\underline{Pressure}}\\[.1cm]
	Pressure sensors typically exploit the change of conductivity due to deformation or expanding of wires and can be integrated in fiber of textiles~\cite{Pervasive_Enokibori_2013,5840}. 
	They are utilised to track footsteps and locations of individuals as well as touch-interaction with the environment~\cite{LocationTracking_Robert_2000}. 
	Installation cost is typically high and requires careful deployment.\\[.2cm]
      \textit{\underline{Ultrasound}}\\[.1cm]
	Ultrasound can indicate relative location of a pair of devices via Time-Of-Flight (TOF)~\cite{5077}.
	Accuracy can be improved via combination with radio frequency~\cite{LocationTracking_Priyantha_2000}.\\[.2cm]
      \textit{\underline{Depth camera}}\\[.1cm]
	Equipped with a depth camera and capable of voice interaction, the Kinect device is able to accurately track gestures of persons~\cite{Pervasive_Ren_2011,Pervasive_Panger_2012} and interaction~\cite{Pervasive_Motta_2013}.\\[.2cm]
	\end{spacing}
	\end{sf}
	\end{scriptsize}
    \end{minipage}
  }
  \fbox{
    \begin{minipage}{.303\textwidth}\begin{scriptsize}\begin{sf}\begin{spacing}{0.9}
      \textbf{Infrastructure-mediated}\\[.2cm]
      Exploitation of alternative sensing modalities which are pre-installed and readily available in environments and therefore minimise installation cost.\\[.2cm]
	\textit{\underline{Resistance; inductive electrical load}}\\[.1cm]
	  Alterations in resistance and inductive electrical load in a residential power supply system can be exploited to detect human interaction in a building~\cite{Pervasive_Patel_2007}.
	  Authors leveraged transients generated by mechanically switched motor loads to detect and classify human interaction from electrical events.\\[.2cm]
	\textit{\underline{Electromagnetic interference (EMI)}}\\[.1cm]
	  Gupta et al. analysed electromagnetic interference (EMI) from switched mode power supplies (SMPS) in order to detect human interaction with electrical systems~\cite{RFSensing_Gupta_2010}.
	It is even possible to detect proximity of the human body towards a Fluorescent Lamp Utilizes from the change in impedance in the EMI structures~\cite{Pervasive_Gupta_2011}\\[.2cm]
	\textit{\underline{Water pressure}}\\[.1cm]
	  Leveraging residential water pipes, the change in water-pressure within the pipe system can be utilised to classify water-related activities and their location in the house (flushing toilet, washing hands, showering,...)~\cite{Pervasive_Campbell_2010,Pervasive_Thomaz_2012,Pervasive_Froehlich_2009}.\\[.2cm]
	\textit{\underline{Gas consumption}}\\[.1cm]
	With a single sensing point, gas use can be identified down to its source (e.g., water heater, furnace, fireplace)~\cite{Pervasive_Cohn_2010b}.
	The authors monitor the gas-flow via a microphone sensor.\\[.2cm]
	\textit{\underline{Electromagnetic noise }}\\[.1cm]
	Using electrostatic discharges from humans touching environmental structure, it is possible to detect locations that have been touched and gestures from electromagnetic noise~\cite{Pervasive_Cohn_2011,Pervasive_Cohn_2012}.
	\end{spacing}
	\end{sf}
	\end{scriptsize}
    \end{minipage}
  }
  \fbox{
    \begin{minipage}{.303\textwidth}\begin{scriptsize}\begin{sf}\begin{spacing}{0.9}
      \textbf{Environmental / Parasitic}\\[.2cm]
       \textit{\underline{Audio}}\\[.1cm]
      Audio can be utilised to identify the location of a phone on room-level and also various in-room (e.g. on table, in drawer) or on-body locations (e.g. pocket)~\cite{ContextAwareness_Kunze_2007}.
      Furthermore, audio-fingerprints can serve as a sense of proximity among devices\cite{Cryptography_Schuerman_2011}.\\[.2cm]
      \textit{\underline{Radio frequency}}\\[.1cm]
      Passive Radar describes a class of radar systems that detect and track objects (vehicles, individuals) by processing reflections from non-cooperative sources of illumination in the environment, such as commercial broadcast and communications signals (HF radio, UHF TV, DAB, DVB, GSM)~\cite{DeviceFreeRecognition_Tan_2005,DeviceFreeRecognition_Colone_2012}.
      In these systems, no dedicated transmitter is involved but the receiver uses third-party transmitters. 
      It then measures the time difference of arrival between Line-of-Sight (LoS) signals and signals reflected from an object.
      By this it is possible to determine the bistatic range of an object and its heading and speed via Doppler Shift and its direction of arrival.
      Expensive systems can operate in ranges of several 100 km but are very expensive. \\
      Recognition of movement is also possible with simpler hardware (WiFi, Sensor nodes, Software-defined-radio) considering the interception of LoS paths between pairs of nodes~\cite{Pervasive_Youssef_2007}.
      In addition, highly accurate localisation was demonstrated by extracting the LoS components among a grid of nodes~\cite{RFSensing_Zhang_2011}. 
      Furthermore, it is possible with similar installations to distinguish activities and gestures (via Doppler fluctuations)~\cite{RFsensing_Pu_2013}, environmental situation~\cite{RFSensing_Ding_2011} as well as attention levels (utilising changes in speed and direction as indicators)~\cite{Pervasive_Shi_2014} and breathing rate~\cite{RFSensing_Patwari_2011b}.\\[.83cm]
      \end{spacing}
   \end{sf}
	\end{scriptsize}
    \end{minipage}
  }
\end{minipage}

 \caption{Overview over various Device-bound and Device-free sensing modalities in the literature}
 \label{figureSensingModalities}
\end{figure*}
The figure summarises popular of these modalities and characterises them for device-bound and device-free (DF) approaches.
Within the device-free techniques, we observe a shift of attention towards the evaluation of environmental, measurable quantities of pre-installed third-party systems which are cheap to use and with increasingly wider physical boundaries.

Researchers have shown remarkable accuracy in tracking activities such as, among others, walking, running, cycling, climbing/descending stairs, sleep states and mobile phone usage~\cite{Pervasive_Berchtold_2010,Pervasive_Bao_2004,Pervasive_Schmidt_2012}.

However, an implicit requirement of these sensing modalities is that the entity or individual to monitor has to cooperate and actually wear the device (device-bound).

In contrast to this, for device-free approaches, the sensing modality need not be worn by the monitored subject.
We can distinguish between classical systems installed particularly for a specific sensing task and systems which are parasitically utilised for sensing but which are originally installed and utilised for other primary purposes. 
Classical device-free systems cover, for instance, video~\cite{Pervasive_Chaquet_2013,ActivityRecognition_Aggarwal_2011}, infrared~\cite{Pervasive_Han_2005,ContextAwareness_Holmquist_2001}, pressure~\cite{LocationTracking_Robert_2000} or ultrasound~\cite{5077,LocationTracking_Priyantha_2000} sensors.
A clear disadvantage of these approaches is their high installation effort. 

This effort can be mitigated by infrastructure-mediated sensing paradigms~\cite{Pervasive_patel_2008,Pervasive_patel_2008b}. 
In general, the approach here is to utilise existing installations, for example, in homes or office buildings, for sensing purposes.
For instance, pressure patterns in residential water pipes might indicate specific activities/usage of inhabitants~\cite{Pervasive_Campbell_2010,Pervasive_Thomaz_2012} or also electromagnetic interference in various electric systems can be utilised to classify activities~\cite{RFSensing_Gupta_2010,Pervasive_Gupta_2011}.
However, these sensing capabilities are limited to indoor application and single buildings.

This limitation is relaxed by systems that utilise environmental sources, such as radio frequency (RF) or audio~\cite{Pervasive_Youssef_2007,ContextAwareness_Kunze_2007}.

In the present survey, we focus on most recent developments in radio-based device-free-recognition.
Such systems monitor changes observed on the RF-channel and analyse them for characteristic patterns.
Changes in the location of objects or movement of individuals causes variation in the radio channel characteristics.
For instance, due to blocked, damped or reflected paths of some of the signals superimposed at a receive node, the absolute signal strength might differ. 
Also, movement might induce Doppler shift in reflected signals and thus lead to changes in the distribution of energy over frequency bands at the receiver.
Figure~\ref{figureRadioEffects} summarises relevant radio effects that can be exploited for environmental awareness from received RF signals.
\begin{figure}
\fbox{
  \begin{minipage}{.945\columnwidth}\begin{scriptsize}\begin{sf}\begin{spacing}{0.9}
  \textbf{Effects on the radio channel}\\[.2cm]
  Radio Frequency (RF) signals are electromagnetic waves, emitted approximately omnidirectional and approximately at speed $c=3\cdot10^8\frac{m}{sec}$ from a transmit antenna.
  At a receiver, all incoming signal components $\zeta_{\mbox{\footnotesize i}}=\Re\left(m(t)e^{j2\pi f_it}\mbox{RSS}_ie^{j(\gamma_i)}\right)$ add up to a received sum signal
  \begin{equation}
\zeta_{\mbox{\footnotesize sum}}=\Re\left(m(t)e^{j2\pi f_ct}\sum_{i=1}^n\mbox{RSS}_ie^{j(\gamma_i+\phi_i)}\right)\label{equationOne}	
\end{equation}
at a center frequency $f_c$. We represent the received signal strength of signal $i$ as RSS$_i$ and its shift in phase from signal generation and due to transmission delay by $\gamma_i$ and $\phi_i$.

In the following, we briefly describe radio effects that are relevant for device-free radio-based recognition systems.
\\[.2cm] 
    \textit{\underline{Multipath propagation}}\\[.1cm]
    Signals might be reflected and scatter at obstacles so that a transmitted signal $\zeta_{\mbox{\footnotesize i}}$ might reach a receiver via varios paths and with different signal delays. \\[.2cm]
    \textit{\underline{Signal fading}}\\[.1cm]
    These incoming copies of an individual signal $\zeta_{\mbox{\footnotesize i}}$ cause constructive and destructive interference at their superimposition at a receiver (fast fading).
    In contrast, slow fading occurs as a result of environmental changes that impact signal propagation (e.g. passing cars, moving trees) 
    \\[.2cm]
    \textit{\underline{Blocking and damping of signals}}\\[.1cm]
    Conditioned on their frequency $f_i$ and the material encountered, a signal c is damped or even blocked by obstacles\\[.2cm]
    \textit{\underline{Doppler shift}}\\[.1cm]
     Relative movement between the transmitter and receiver incurs a change in frequency of a signal $\zeta_{\mbox{\footnotesize i}}$. 
      This Doppler shift $f_d$ is conditioned on the relative speed $v_i$ between transmitter and receiver, the frequency $f_i$ and the angle $\alpha_i$ of the movement direction between transmitter and receiver:
     $f_d=\frac{v_i}{\lambda}\cdot\cos(\alpha_i)$
     \\[.2cm]
     \textit{\underline{Path loss}}\\[.1cm]
     The signal strength of an RF-signal reduces with distance. 
     A straight forward calculation of this path loss can be calculated by the Friis Free space equation~\cite{CommunicationTechnology_Friis_2946}
    \begin{equation}
      P_{RX}=P_{TX}\cdot \left(\frac{\lambda_i}{2\pi d_i}\right)^\tau\cdot G_{TX}\cdot G_{RX}     
    \end{equation}
     Here, $P_{TX}$ describes the transmit signal strength, $G_{TX}, G_{RX}$ represent the antenna gain at transmit and receive devices, $\lambda_i=\frac{c}{f_i}$ describes the wave length and $d_i$ is the distance traversed.
    The path-loss exponent $\tau$ differs with the environment and typically takes values between $2$ and $5$. 
      \end{spacing}
     \end{sf}
	\end{scriptsize}
  \end{minipage}
  }
\caption{Summary of some radio effects that can be exploited for RF-based Device-Free recognition}
\label{figureRadioEffects}
\end{figure}

An early example of a system utilising WiFi signals for the localisation of a receive device is the RADAR system that employed signal strength and signal-to-noise-ratio (SNR) from WiFi~\cite{Pervasive_Bahl_2000}.
Other implementations utilised GSM for localisation by employing signal strength readings from the active set~\cite{Pervasive_Otsason_2005,Pervasive_Varshavsky_2007} or signal strength from a set of FM base stations~\cite{Pervasive_Krumm_2003,Pervasive_Youssef_2005}.
Frequently, these approaches require the creation of a received signal strength (RSS) fingerprint map~\cite{Pervasive_Jiang_2012,Pervasive_Pulkkinen_2012,DeviceFreeRecognition_Aly_2013}, but also real-time on-line localisation that does not require a fingerprinting map is feasible~\cite{Pervasive_Schougaard_2012,Pervasive_Wang_2012,Pervasive_Chen_2012,Pervasive_Chen_2012b,Pervasive_Chen_2012-2}. 
The latter approaches combine, for instance, dead reckoning methods with characteristic, crowd identified, waypoints for accurate relative localisation.
These systems are device-bound and can reach high accuracy of about 1~meter~\cite{RFSensing_Sen_2012}.

\begin{figure*}
\includegraphics[width=\textwidth]{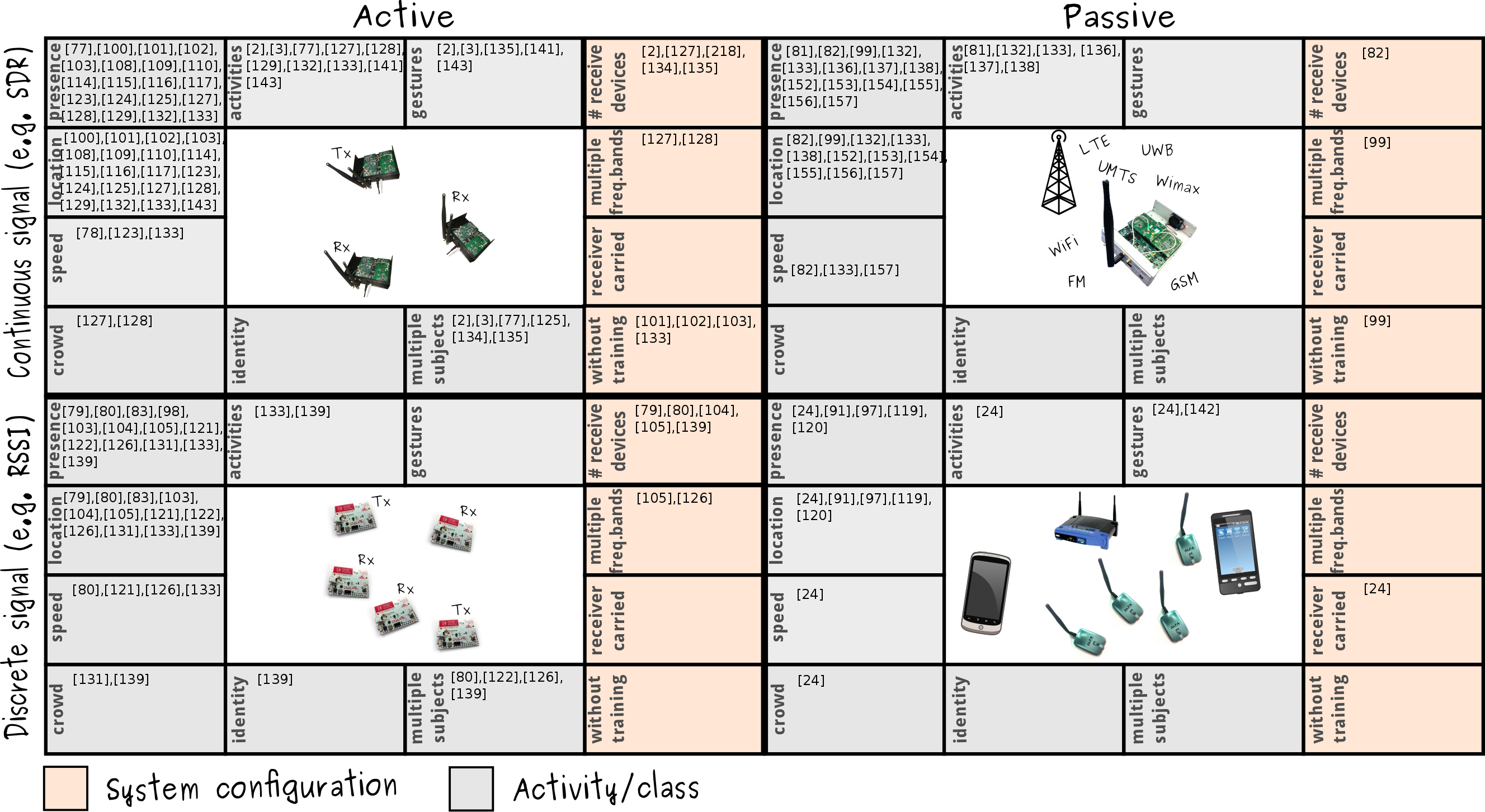}
\caption{RF-based device-free activity recognition systems and their recognition capabilities and system configuration considered. The figure groups related the corresponding reference to reach system under the respective class. }
\label{figureDFARClasses}
\end{figure*}

For device-free approaches, on the other hand, the monitored entity is not equipped with any transmit or receive device~\cite{Pervasive_Youssef_2007}.
We distinguish between four classes of such recognition systems conditioned on their hardware configuration (cf. figure~\ref{figureDFARClasses}).
These systems can be grouped into active and passive approaches conditioned on the presence of an active transmitter~\cite{Pervasive_Scholz_2011b}.
Active systems control both, transmit and receive hardware while passive systems only utilise receive devices. 
Most current systems are active such that both, the receiver and the transmitter are under the control of the system.
Generally, the classification accuracy of an RF-based device-free recognition system suffers when the transmitter is third-party controlled.

Many early studies utilise continuous signals captured by Software-Defined Radio (SDR) devices for their more accurate and complete access to the radio channel. 
These systems can exploit continuous signals received on the wireless channel and sampled at a high frequency, which enables the utilisation of frequency domain features.

In contrast, consumer devices seldom feature SDR-capabilities. 
On such devices, frequently, the Received Signal Strength Indicator (RSSI) is exploited as an indicator for surrounding activities and situations.

Figure~\ref{figureDFARClasses} indicates research achievements demonstrated for the respective classes and system configurations by various groups.
Most results have yet been achieved for active, continuous signal based systems. 
On the contrary, passive RSSI-based systems are only recently considered.
In addition, most work considers the recognition or localisation of individuals (presence, location). 
For continuous signal-based systems also more complex cases like activities have been considered. 
More complex system configurations or classes are to-day less frequently investigated and partly also constitute open research questions.

The following sections detail the research conducted in these fields in more detail and also cover comparative measures like accuracy of recognition.

\subsection{Localisation}
Device-free RF-based recognition was first investigated for the task of localisation or tracking of an individual.
Youssef defines this approach as Device-Free Localisation (DFL) in~\cite{Pervasive_Youssef_2007} to localise or track a person using RF-Signals while the entity monitored is not required to carry an active transmitter or receiver.

In the following, we distinguish between preliminary studies considering basic impacts of presence and movement on a received radio signal, radio tomographic imaging approaches, RF-fingerprinting methods, anomaly detection methods and approaches that isolate direct links among nodes in order to analyse their fluctuation.

\subsubsection{Impact of presence and movement}
Youssef et al. analysed the impact of presence on a received radio signal and defined three tasks for DFL: detection of presence, tracking of persons and predicting identity of individuals~\cite{Pervasive_Youssef_2007}.
For the mere detection of presence, they analysed the moving variance and moving average of the time-domain signal strength of RSSI values from transmitting and receiving pairs of WiFi devices (access points (AP) and mobile terminals).
Classification accuracy reached up to $1.0$ for some configurations.
In order to track individuals they proposed the use of a passive radio map (see section~\ref{sectionRFFingerprinting}).

Kosba et al. presented in~\cite{Pervasive_Kosba_2011} a similar system to detect motion from RF-readings of standard WiFi hardware.
Their system utilises a short offline training phase in which no movement and activity is assumed as a baseline. 
Then, anomaly detection is employed in order to detect changes from that baseline.
The authors considered mean or variance-related features and concluded that the variance is better suited to detect changes in the RSSI.
In contrast to the works of Zhang and others, this system does not require WiFi nodes to be located in an exactly defined grid with fixed node distances.
Consequently, localisation is not possible but mere detection of presence. 

Also, Lee et al. consider the utilisation of RSSI fluctuation from pairs of communicating TelosB nodes for intrusion detection~\cite{RFSensing_Lee_2010}.
In five distinct environments (outdoor and indoor) they reported changes in the mean and standard deviation of absolute RSSI values. 

Utilising a passive, FM-radio based system with SDR devices, Popleteev indicated that frequency diversity can help to improve localisation accuracy of RF-based systems~\cite{DeviceFreeRecognition_Popleteev_2013}.
In particular, he considered a person located at 5 different locations inside a room and predicted the location with a standard k-nearest neighbour approach. 
In addition, the author pointed out that the classification accuracy of the system would deteriorate when the system is trained on one day but classification is conducted on another day.

Lieckfeldt and others considered the impact of the presence of a single individual on the received signal strength observed by an RFID reader in a 2m$\times$2m area equipped with 69 passive RFID tags~\cite{DeviceFreeRecognition_Lieckfeldt_2009}.
Their system utilised a two-staged approach in which first the RSSI fluctuation without presence was recorded and later, presence was detected via the observed changes in the signal strength from the set of tags.
The authors observed that the backward link is more expressive for the recognition of presence than the forward link from the reader. 
In addition they considered different orientations of the monitored individual in order to arrive at more general results.

\subsubsection{Radio tomographic imaging}
Tomography desribes the visualisation of objects via a penetrating wave.
An image is then created by analysing the received wave or its reflections from objects.
A detailed introduction to obstacle mapping based on wireless measurements is given in~\cite{RFSensing_Mostofi_2013b,RFSensing_Mostofi_2011}.
Radio tomography was, for instance, exploited by Wilson et al. in order to locate persons through walls in a room~\cite{RFSensing_Wilson_2009}.
In their system, they exploit variance on the RSSI at 34 nodes that circle an area in order to locate movement inside that area.
Nodes in their system implement a simple token-passing protocol to synchronise successive transmissions of nodes.
these transmitted signals are received and analysed by the other nodes in order to generate the tomographic image by heavily relying on Kalman filters.
They were able to distinguish a vacant area from the area with a person standing and a person moving. 
In addition, it was possible to identify the location of objects and to track the path taken by a person walking at moderate speed.
An individual image is taken over windows of 10 seconds each.
By utilising the two-way RSSI fluctuations among nodes, an average localisation error of 0.5 meters was reached~\cite{RFSensing_Wilson_2010}.

\begin{figure}
 \centering
 \includegraphics[width=\columnwidth]{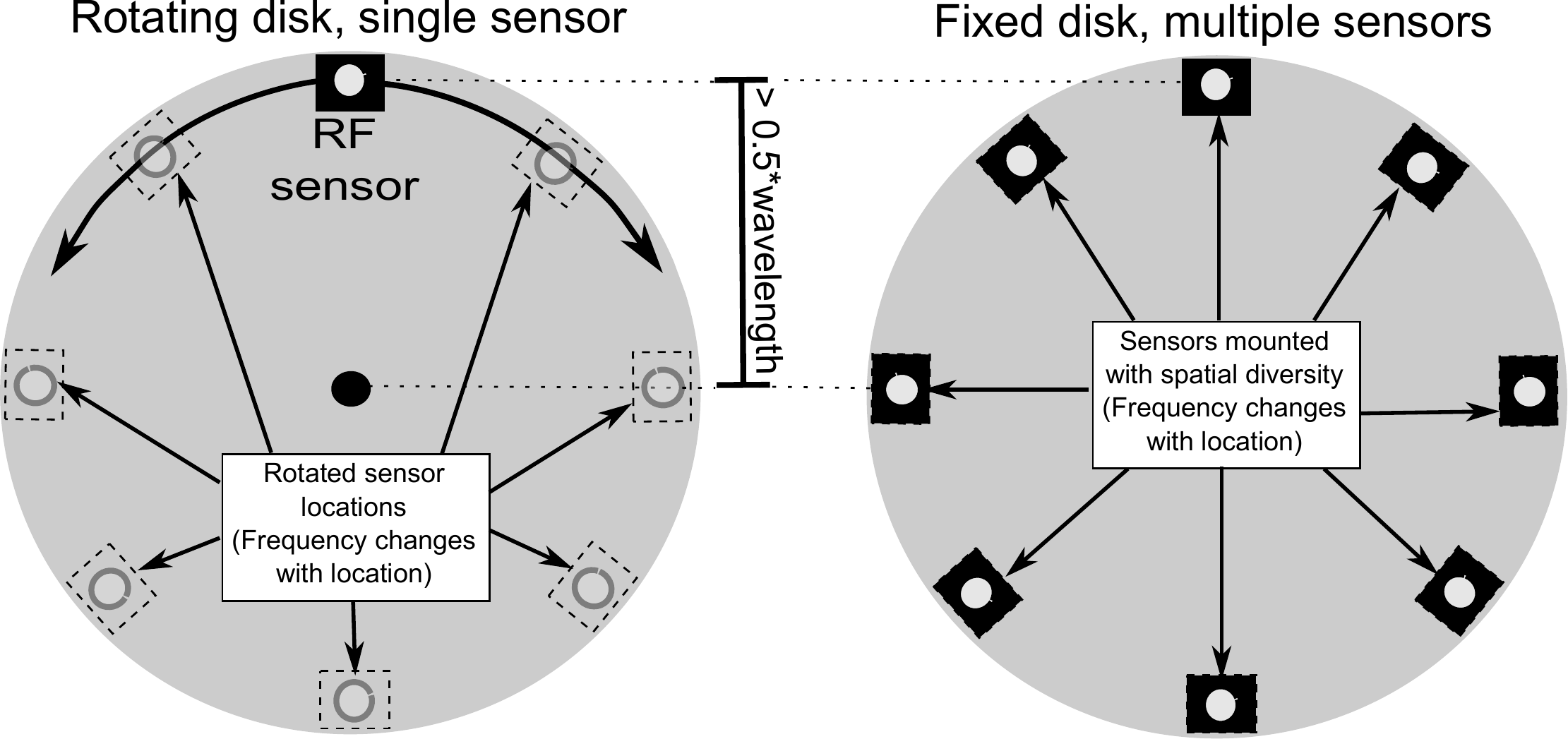}
 \caption{Illustration of the utilisation of an RF-sensors exploiting spatial diversity via a rotating disk or multi-node instrumentation~\cite{RFSensing_Bocca_2013}}
 \label{figureRotatingNodes}
\end{figure}

It was reported in~\cite{RFSensing_Bocca_2013} that the localisation accuracy of such a system can be greatly improved by slightly changing the location of sensors, thus exploiting physical diversity.
The authors present a system in which nodes are attached to disks equipped with motors in their center for rotation as depicted in figure~\ref{figureRotatingNodes}. 
With this setting it is possible to iteratively learn a best configuration (physical location) of nodes similar to, for instance, iterative beamforming approaches that try to lock several radio signals on the optimal relative phase offset~\cite{4022,DistributedBeamforming_Quitin_2013}.

Wagner et al. implemented a radio tomographic imaging system with passive RFID nodes instead of sensor nodes. 
Implementing generally the same approach as described above, they could achieve good localisation performance with their system. 
However, they had to implement a suitable scheduling of the probabilistically scattered transmissions of nodes due to the less controllable behaviour of passive RFID nodes~\cite{DeviceFreeRecognition_Wagner_2012}.
In later implementations, they improved their system to allow on-line tracking~\cite{DeviceFreeRecognition_Wagner_2013-2} and a faster iterative clustering approach to further speed up the time to the first image generated~\cite{DeviceFreeRecognition_Wagner_2013}. 
This image is then of rather low accuracy but is iteratively improved in later steps of the algorithm. 
With this approach, it was possible to achieve a localisation error of about 1.4m after only one second and reach a localisation error of 0.5m after a total of about seven seconds in a 3.5m$^2$ area.

Utilising moving transmit and receive nodes and compressive sensing theory~\cite{CompressiveSensing_Candes_2006,CompressiveSensing_Donoho_2006,CompressiveSensing_Needell_2010} it is possible to greatly reduce the number of nodes required.
For instance, Gonzalez-Ruiz et al. consider mobile robotic nodes that mount transmit and receive devices and circle the monitored target in order to generate the tomographic image~\cite{RFSensing_Gonzalez_2013}.
In particular, they required only two moving robots attached with rotating angular antennas in order to accurately detect objects in the monitored area.  
Each robot takes new measurements every two centimeters. 
Overall, after about 10 seconds a single image can be taken.
They detail their implemented framework in ~\cite{RFSensing_Gonzalez_2014} and the theoretical framework for the mapping of obstacles, including occluded ones, in a robotic cooperative network, based on a small number of wireless channel measurements in~\cite{RFSensing_Mostofi_2013}.
 
\subsubsection{Machine learning}
Instead of generating radio-tomographic images, which is an accurate but comparatively slow procedure, also general Machine Learning approaches can be employed for RF-based localisation.
For instance, Wagner et al. investigate the localisation in a passive RFID setting utilising multi layered perceptrons for training-based device-free user localization~\cite{DeviceFreeRecognition_Wagner_2012-2}.
In particular, the authors utilised a three-layer neural network that takes the a series of measurements as input vector and provides a tuple as output defining a two-dimensional user location.
Localisation error achieved could be kept below 0.5 meters in a 3m$\times$3m square area.

\subsubsection{RF-Fingerprinting}\label{sectionRFFingerprinting}
A common approach to RF-based localisation is the construction of radio strength maps.
In device-based systems, RSS at various locations is tracked and used as a map together with access point IDs~\cite{Pervasive_Jeon_2013}.
With this information, location is later estimated from life measurements.
Such radio maps may also be deployed with device-free systems in which the RSSI fluctuations in the presence of a person not equipped with a transmit or receive device are captured. 
Youssef et al. present such a localisation system in~\cite{Pervasive_Youssef_2007}.
They report that the RSSI is more stable over night when no people are around so that this is the best time to create an RSSI fingerprint map. 
In a system with two transmit and two receive WiFi devices monitoring the RSSI in infrastructure mode from beacons sent roughly every 100ms they have been able to accurately predict and trac location of a single person in an indoor location.
Later, they improved their approach using less nodes~\cite{Pervasive_Seifeldin_2009}.
This was possible by employing a Bayesian inference algorithm.
All these experiments have been conducted under Line-of-Sight (LoS) conditions.
A major drawback has been the time-consuming manual generation of the fingerprint maps, however, with current systems, also automated generation of RSSI fingerprints on laptop-class computers is possible~\cite{Pervasive_Seifeldin_2013} 
 
\subsubsection{Geometric models and estimation techniques}
Finally, in systems where the relative location of nodes that transmit and receive signals is exactly known, the geometry and layout of the instrumentation can be exploited. 
Zhang et al. employed a grid of nodes in order to localise individuals from device-free WiFi readings~\cite{DeviceFreeRecognition_Zhang_2007}.
\begin{figure}
 \centering
 \includegraphics[width=\columnwidth]{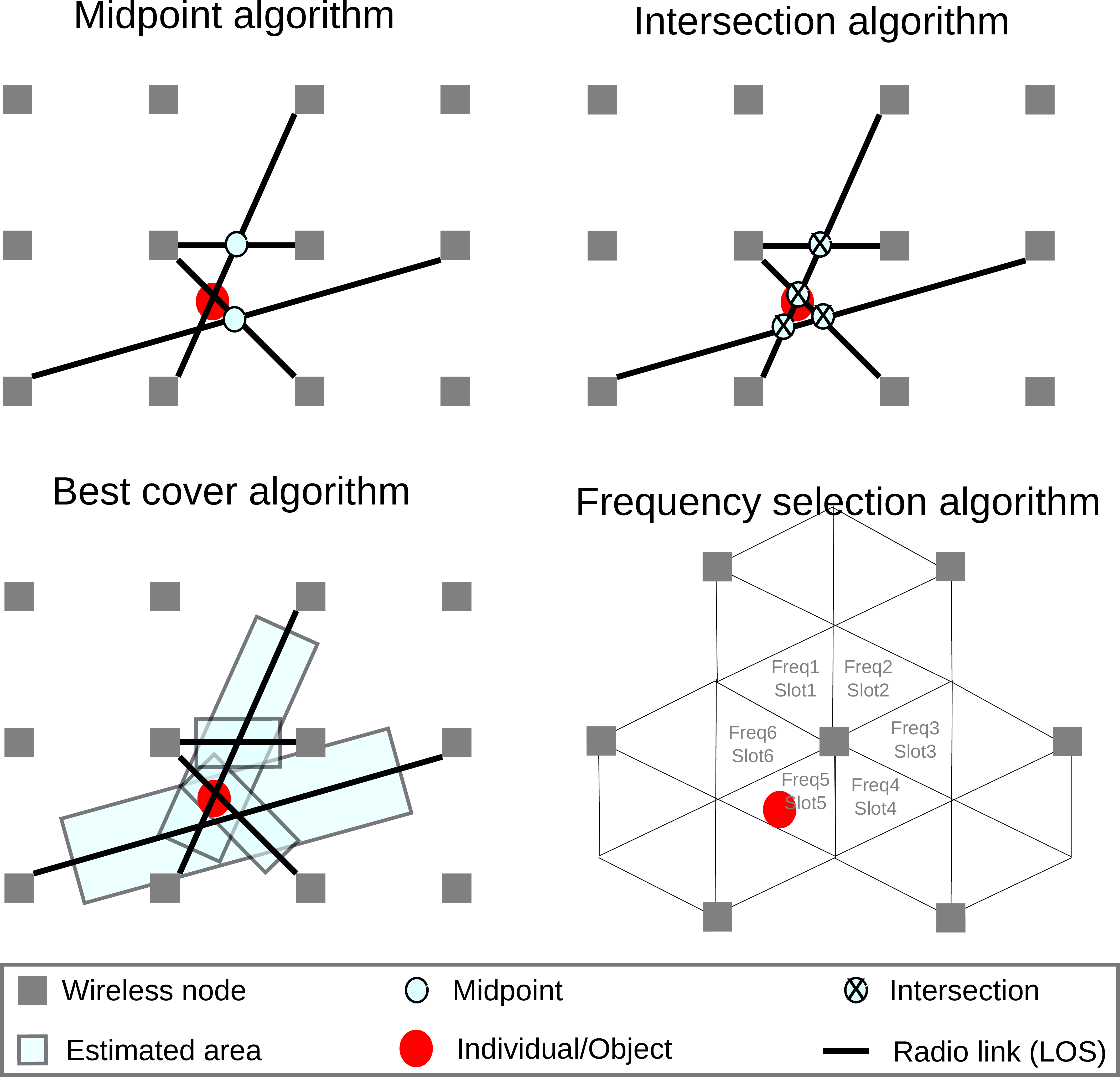}
 \caption{Device-free RSSI-based localisation of objects via four distinct algorithms~\cite{DeviceFreeRecognition_Zhang_2007,RFSensing_Zhang_2011}}
 \label{figureRSSIBasedLocalisationAlgorithms}
\end{figure}
They proposed a straightforward theoretic model to describe signal fluctuation induced by passive objects and verified their findings in a case study with ceiling mounted MICA2 sensor nodes transmitting with 0dBm at 870MHz.
The three algorithms proposed (Midpoint, Intersection, Best cover) all require an initial training phase in which the RF fluctuation is monitored in a stable state with no interference through individuals (cf. Frequency Selection Algorithm in figure~\ref{figureRSSIBasedLocalisationAlgorithms}). 
All algorithms utilise knowledge about the relative location of nodes and exploit RF-signal strength fluctuation on direct links.
From this, center locations on the direct links, Intersections of direct links or 0.5$\times$0.5m$^2$ areas on the direct links are utilised in order to predict the location of activity. 
Best results have been achieved with the consideration of overlapping areas.
The optimum distance among two nodes in the grid has been experimentally derived as 2 meters.
With this configuration, a single person moving slowly (0.5 m/s) along a straight line has been tracked with an accuracy of below 1m and two persons with an accuracy of below 2m. 
With additional clustering of nodes, the accuracy for the tracking of multiple persons could be further improved to slightly more than 1m~\cite{RFSensing_Zhang_2009}.
Also, the transmission power was demonstrated to impact the tracking accuracy and lower transmission powers of $-6$ to $-11$~dBm have been observed to show more dynamic values for short node distances.
The system was shown to be real-time capable in~\cite{RFSensing_Zhang_2011}.
By clustering the measurement area into several, frequency-separated cells, spanned by three nodes each, the authors could isolate interference from neighbouring nodes and also speed up the computation (cf. figure~\ref{figureRSSIBasedLocalisationAlgorithms}).

Utilising passive RFID transponders, Lieckfeldt et al. exploited device-free Localization in recent years \cite{RFSensing_Lieckfeldt_2009,RFSensing_Lieckfeldt_2009b}.
The authors propose a physical model that depicts the effect of relative position of subjects on the signal strength. 
They propose estimators for user localization, based, for instance, on maximum likelihood and geometric methods, such as nearest intersection points.
While the geometric approaches suffer from a low accuracy, the estimation based methods are characterised by a high computational complexity.


A straightforward approach to localisation based on RSSI fluctuation is the consideration of the interception of LoS paths in a grid of nodes. 
A first step in this direction was taken by Patwari et al. who derived a statistical model for the RSS variance as a function of the location of a single individual~\cite{RFSensing_Patwari_2011}.
They could show that reflection causes the RSS variance contours to be shaped approximately like Cassini ovals. 
They also considered the simultaneous localisation of multiple individuals at the same time and argue that their model could be extended to cover multiple individuals.
This was later demonstrated to be feasible in an actual system instrumentation by Zhang and others~\cite{Pervasive_Zhang_2012}.
The authors isolate the LoS path by extracting phase information from the differences in the RSS on various frequency spectrums at distributed nodes.
Their experimental system is with this approach able to simultaneously and continuously localise up to 5 persons in a changing environment with an accuracy of 1~meter.

\subsection{Recognition of activities}
Not only static location but also activities, gestures or situation in proximity of a receive antenna can be distinguished from signal fluctuation over time. 
For RF-based activity recognition, a higher sampling frequency is required than for mere localisation or tracking. 
Depending on the specific application, sampling rates between 4Hz and 70Hz are utilised.
Consequently, methods such as tomographic imaging are too slow to achieve reasonable accuracy here.
Furthermore, as location is not the main interest, geometric models and RF-fingerprinting are not employed.
Especially the latter captures static situations and can therefore not be applied for the recognition of dynamic changes over a time window.

  Instead, machine learning techniques are frequently applied to analyse fluctuation in signal strength  measurements over time.
  In addition to RSS, also movement-indicating features such as frequency-domain features or Doppler shift are exploited. 

\subsubsection{Machine learning and estimation}
In their seminal work, Patwari et al. report that they are able to detect the breathing rate of a single individual by analysing the RSS fluctuation in received packets from 20 nodes surrounding the subject~\cite{RFSensing_Patwari_2011b}.
Via maximum likelihood estimation, they were able to estimate the breathing rate with a Root-Mean-Square-Error (RMSE) of 0.3 breaths per minute.
Their system consists of Telos~B nodes transmitting every 240ms on a center frequency of 2.48 GHz, which translates to an overall packet transmission rate of about 4.16Hz.
Prediction was taken after a 10 second to 60 second measurement period. 
Best results could be achieved with 25 to 40 seconds whereas longer observation periods did not further improve the accuracy significantly.
Naturally, the accuracy achieved was dependent on the number of nodes that participated. 
While a single node pair could not achieve usable results, already with 7 network nodes, an RMSE breathing rate error of only about 1.0 was observed.
They could further show that the links with low average RSS are most significant for the detection of breathing rate.

\begin{figure}
\centering
  \includegraphics[height=0.9\columnwidth, angle=90]{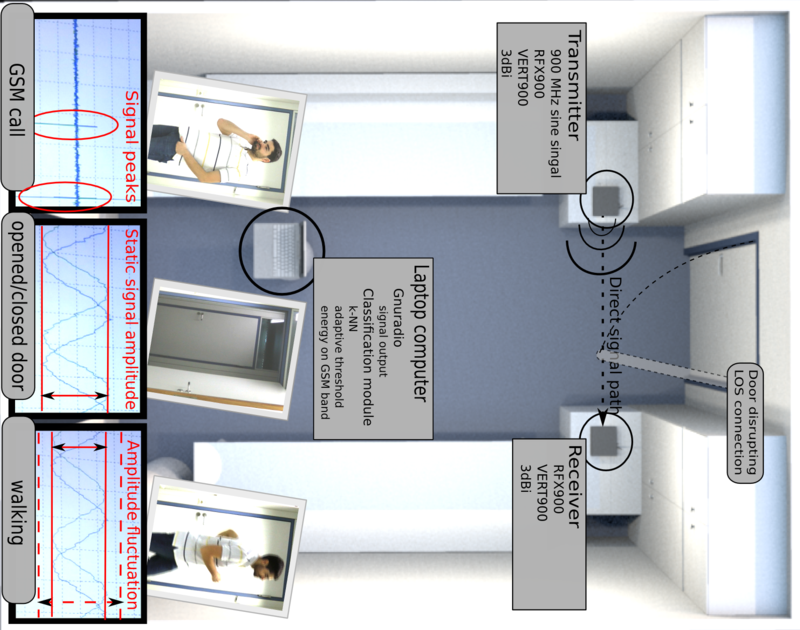}
  \caption{Recognition of three well separated classes in the SenseWaves system~\cite{Pervasive_Scholz_2011}}
\label{figureSystem}
\end{figure}
With standard machine learning approaches (e.g. k-nearest neighbour, decision tree, Bayes, support vector machines), it is possible to extract further information on environmental situation from RSS fluctuation.
In preliminary studies, Reschke, Scholl, Sigg and others demonstrated the detection of opened or closed doors, presence and crowd size with an accuracy of 0.6 to 0.7~\cite{4036,ContextAwareness_Sigg_2011,Pervasive_Scholz_2011,OrganicComputing_Sigg_2011} (figure~\ref{figureSystem} illustrates the SenseWaves recognition system for the distinction of three fairly separated classes).

The authors utilised USRP Software defined radio devices (SDR)\footnote{http://www.ettus.com} from which one constantly transmits a signal at frequencies between 900MHz to 2.4GHz that is read and analysed by other nodes.
The SDR devices allow high sampling rates of the observed signal. 
In their system, the authors employ sampling rates of 40Hz from a continuous signal transmitted by one node.
No specific relative placement of nodes was required so that the system qualifies for ad-hoc deployment. 
For recognition, simple time-domain RSS features such as the Root of the Mean Squared (RMS), Average Magnitude Squared (AMS), Signal-to-Noise Ratio (SNR)~\cite{4036,ContextAwareness_Sigg_2011,OrganicComputing_Sigg_2011}, signal amplitude, signal peaks in a defined time period and the number of large deltas in successive signal peaks~\cite{Pervasive_Scholz_2011} have been utilised.
Also, the consideration of crowd size extends the often followed single-individual sensing approach~\cite{RFsensing_Xu_2013}.
The author's learning approach is able to predict the count of up to 10 stationary or moving individuals.

\begin{figure}
     \begin{center}
     \includegraphics[width=\columnwidth]{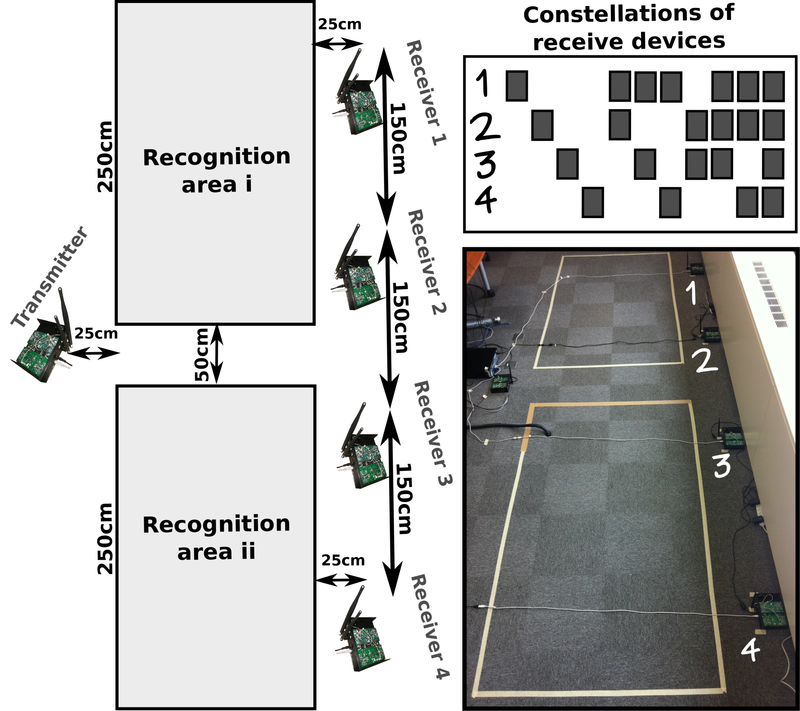}
     \caption{Constellations of 1, 2, 3 or all receivers for the simultaneous recognition of activities from multiple subjects~\cite{DeviceFreeRecognition_Sigg_2013}}
     \label{figureEnvironment}
          \end{center}
\end{figure}

Later, with the consideration of additional and also frequency domain features, recognition accuracy was further improved~\cite{Pervasive_Sigg_2012, Pervasive_Sigg_2013}.
In addition, the authors compared several device-free recognition techniques and also accelerometer-based recognition with the result that the active and passive device-free and continuous signal based systems could score similar results as accelerometer-based recognition systems.
The authors also reported that some features such as the variance are robust against static environmental changes for the detection of dynamic activities, such as walking or crawling.
In addition, it was possible to distinguish activities conducted by multiple persons simultaneously in an active SDR-based system.
With two persons conducting activities at two locations and four receive devices, the authors trained the classifiers on the combined features and could distinguish 25 cases with high accuracy~\cite{DeviceFreeRecognition_Sigg_2013} (cf. figure~\ref{figureEnvironment}).
Later, the recognition of gestures in the proximity of a receive antenna was reported with a similar approach~\cite{Pervasive_Sigg_2014b}. 

In a related work with an SDR-based but passive system, Shi et al. exploited signals from a nearby FM radio station for the detection of activities.
Their method also exploits machine learning approaches but relies more on frequency domain features.
In addition, their sampling rate is lower with about 2Hz and a sampling window of 0.5 seconds~\cite{Pervasive_Shi_2012, Pervasive_Shi_2012b, DeviceFreeRecognition_Shi_2013}. 
However, the accuracy achieved is comparable to the above active systems.

Another approach utilising RSSI information from sensor nodes in an active RSSI-based system was presented in~\cite{Pervasive_Scholz_2013}.
The authors place eight 802.15.4 nodes that transmit at 2.4 GHz in a 20m$^2$ office room.
The nodes were placed at various heights from 30cm to 1.4m.
With this setting and only mean and variance as features, the authors could distinguish seven different classes at an accuracy that exceeded the accuracy achieved by an accelerometer attached to the subject for comparison. 
They reported that their 3D topology helps to distinguish activities and that there are indications that discrimination of subjects might also be possible.

Very recently, Sigg et al. investigated the distinction of gestures and situations in a passive device-free system with only one off-the-shelf (smartphone) receiver~\cite{Pervasive_Sigg_2014,RFSensing_Sigg_2014}.
\begin{figure*}
 \includegraphics[width=\textwidth]{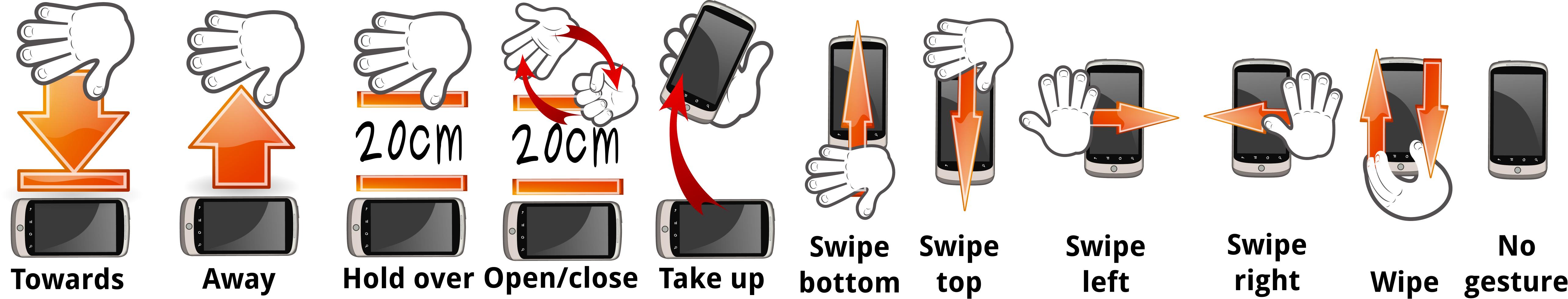}
 \caption{Gestures recognised via RSSI fluctuation on an off-the-shelf mobile phone~\cite{RFSensing_Sigg_2014}}
 \label{figureGestures}
\end{figure*}
They observed that 10 RSSI packets per second could be expected in urban places and that these are sufficient to distinguish between simple classes and also hand gestures in proximity of the receiver.
Although their accuracy reached was lower than for the active RSSI-based system reported above, it was clearly above random guess.
In addition they could distinguish 11 gestures performed in close proximity of the phone.

\subsubsection{Doppler Shift}
When an object reflecting a signal wave is in motion, this causes Doppler Shift. 
The direction and speed of the movement conditions the strength and nature of this frequency shift.
Pu and others showed that simultaneous detection of gestures from multiple individuals is possible by utilising multi-antenna nodes and micro Doppler fluctuations~\cite{RFsensing_Pu_2013,RFsensing_Kim_2009}.
They utilise a USRP SDR multi antenna receiver and one or more single antenna transmitters distributed in the environment to distinguish between a set of 9 gestures with an average accuracy of 0.94. 
Their active device-free system exploits a MIMO receiver in order to recognise gestures from different persons present at the same time. 
By leveraging a preamble gesture pattern, the receiver estimates the MIMO channel that maximises the reflections of the desired user.

A main challenge was for them that the Doppler shift from human movement was several magnitudes smaller than the bandwidth of the signal employed.
The authors therefore proposed to transform the received signal into several narrowband pulses which are then analysed for possible Doppler fluctuation.
The group discussed application possibilities of their system in~\cite{RFSensing_Kellog_2014}.

In a related system, Adib and Katabi employ MIMO interference nulling and combine samples taken over time to achieve a similar result while compensating for the missing spatial diversity in a single-antenna receiver system~\cite{Pervasive_Adib_2013}.
In their system, they leverage standard WiFi hardware at 2.4GHz.

Later, this work was extended to 3D motion tracking by utlising three or more directional receive antennas in exactly defined relative orientation~\cite{RFSensing_Adib_2014}. 
In particular, the system is able to track the center of a human body with an error below 21cm in any direction and can also detect movement of body parts and directions of a pointing body part, such as a hand. 
This localisation is possible through time-of-flight estimation and triangulisation.
Higher accuracy of this estimation is granted by utilising frequency modulated carrier waves (sending a signal that changes linearly in frequency with time) over a bandwidth of 1.69GHz.
Impact of static objects could be mitigated by subtracting successive sample means whereas noise was filtered by its speed of changes in energy over frequency bands. 

\section{Toward Cognitive Activity Recognition}
\label{sectionCognitive}
Physical activity tracking came a long way, from dedicated sensing devices in lab settings to consumer applications embedded in wearable appliances (e.g. Fitbit, Jawbone UP) and even dedicated human motion tracking co-processors in smart phones (e.g. M7 in the iPhone 5s). 
Now we are seeing the first end consumer devices that start exploring our physiological signals (heart rate, blood oxygen level etc.) and our sleep performance. 

The next logical step is the tracking of cognitive activities:
attention, recall, cognitive load and finally learning and decision making. We explore in this section which sensor modalities seem to have the most merit and then tackle a very specific type of cognitive task, namely reading. 
We discuss why reading is a good choice to start with and how we tracking can be extended towards other cognitive activities
\cite{kunze2013activity}.

\subsection{Importance of Eye Gaze}
The most obvious way to track cognitive tasks is to monitor the brain directly. 
Although this approach sounds promising, there are a lot of practical problems with direct brain monitoring. 
Either the methods are very obtrusive (e.g. fMIR) or they have problems with noise, movement artifacts and are not easy to wear during everyday life.

As intermediate technology, eye movement tracking seems to be the most promising. 
As it can be easily monitored either using optical eye tracking or electrooculography (electrodes placed close to the eye). 
Also, eye movements are closely correlated to cognitive activities and states.
From simple blink frequency analysis that can tell you about the fatigue of a user to expertise analysis for complex visualizations. 

\subsection{Quantifying Reading Habits}
In this section, we focus on tracking reading as a cognitive activity. 
There are two main reasons. 
First, reading is a ubiquitous task, performed everyday crucial to our learning and knowledge acquisition. 
Although there are very detailed studies of reading activities in the lab, there are very few in-situ studies about reading behavior in real life. 
Second, we believe "reading" can become to cognitive activity tracking what "walking" and locomotion analysis became for the physical task tracking. 
Reading analogous to Walking is easy to define and includes repetitive movements with distinct frequencies. 
This should make the task of spotting it easier, while preventing the definition problem. 
Take "focusing" or "paying attention" as an example, for spotting cognitive activities depends highly on how you define them. 

\begin{figure}
    \includegraphics[width=\columnwidth]{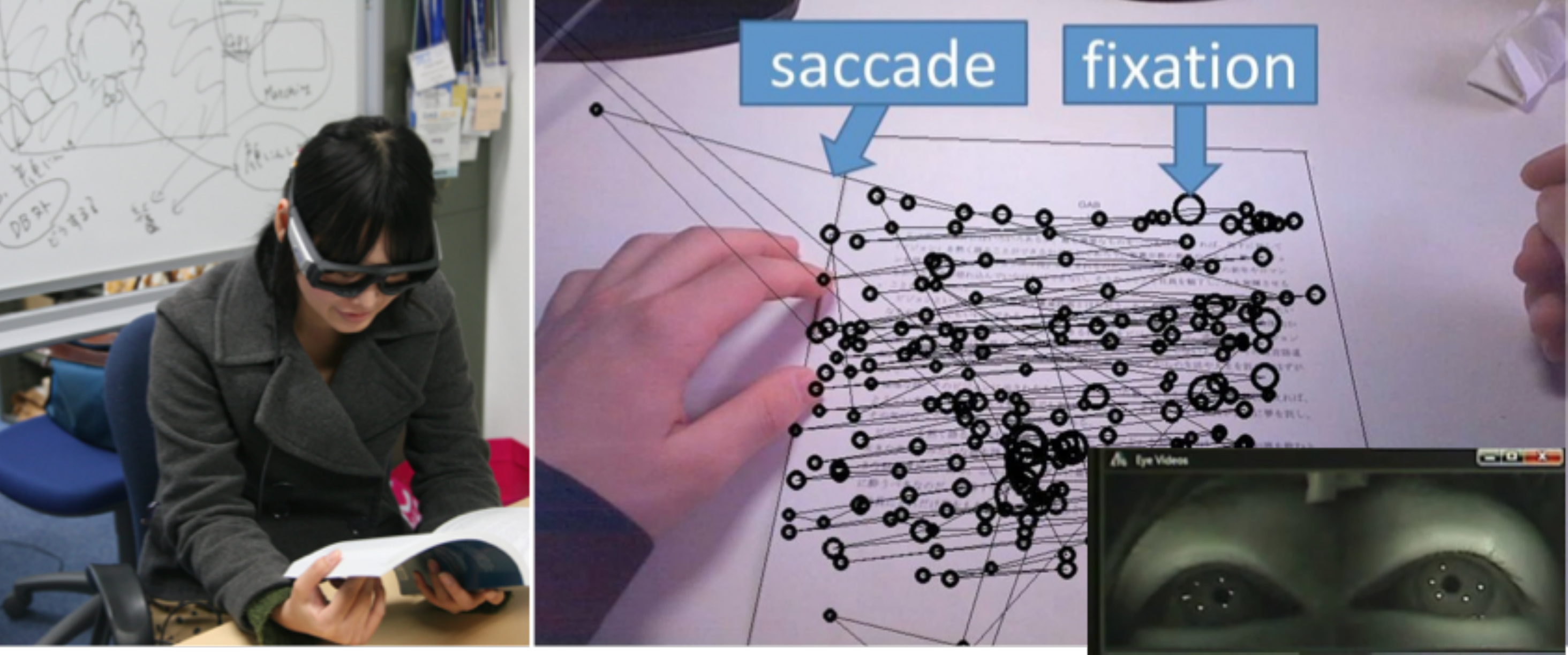}
   \caption{User reading a document with head-mounted eye tracker.}
   \label{eyetracker}
\end{figure}

To track reading habits we evaluated a couple of technologies (e.g. EEG, eye tracking, motion sensors, egocentric cameras) and found mobile eyetrackers are so far the best suited for the task (see Fig.~\ref{eyetracker} for an exemplary setting with a person wearing a mobile eyetracker). 
Our analysis goes from simple word count over reading material inference to trying to assess reading comprehension.

\subsubsection{Word Count and Reading Speed}
It is possible to implement a wordometer using optical mobile eye tracking~\cite{icdar1}. 
The number of words a user reads can be counted by recognizing reading, counting line breaks and then approximating the words read. 
Current implementations works with an error rate of around 6-11\% for 10 users over 20 documents with sizes ranging from 150-680 words.

The recognition process works as follows. 
First, reading is recognized by a support vector machine using fixation and saccade features~\cite{icdar1}. 
Afterwards, there are several ways to estimate reading volume: using time only, detecting a line break (long saccade back towards a new line) to estimate lines, based on the lines or the word count.
The latter method works better (5-15\% lower error rate).

Reading volume in itself is associated with an increase in vocabulary and there are strong correlations between size of vocabulary and language skill.
However, more interestingly, reading volume seems also an indicator for higher general knowledge~\cite{cunningham2001reading}. 
In itself, reading volume is therefore already interesting information. 
Yet, it can also enable novel applications, like annotating books with the amount of reading a user did (and at which pages) to give feedback to authors.

\subsubsection{Document Type Classification}
\begin{figure*}
    \centering
    \includegraphics[width=7.0in]{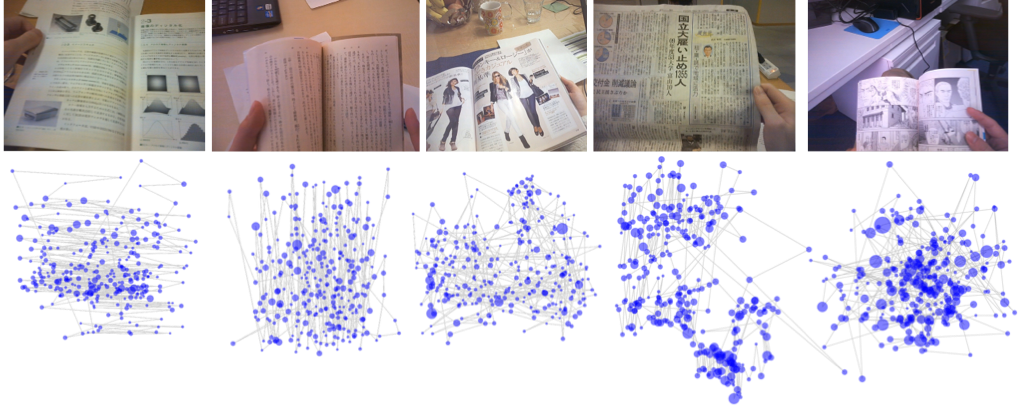}
   \caption{Examples for different eye gaze patterns for varying document types (Texbook, novel, magazine, newspaper, manga)}
   \label{doctype}
\end{figure*}

Using also a mobile eyetracker, it is possible to tell which documents a user reads. 
In figure~\ref{doctype}, exemplary eye-gaze patterns are displayed for various document types (comic book, text book, magazine etc.). 
In an experiment with 10 users reading 5 different document types for 10 minutes in 5 different environments (e.g. office, coffee shop) an accuracy of 78\% for around 1~minute windows are achieved independent of the user and 98\% for the user dependent case.
As long as the document layout is sufficiently unique, information about the document is also contained in the eye movement~\cite{kunze2013know}.

This raises the interesting question, if given a particular goal, there are optimal eye gaze patterns for reading a particular document. 
If this were the case, we could store the optimal eye gaze pattern and adjust the document accordingly if the user deviates from that pattern. 

\subsubsection{Toward Reading Comprehension}
In the same line of research yet even more difficult, researchers assess whether it is possible to estimate expertise level from eye gaze. 

So far, the results are ambiguous regarding the estimation of reading comprehension. 
Although there is a clear correlation between a couple of eye gaze features and the comprehension of the reader, the data seems noisy making a good inference difficult. 
We can detect difficult words by using fixation counts for individual users, yet so far it was not possible to determine reading comprehension directly \cite{kunze2013activity}.
Difficult word detection is based on fixation count.
Difficult words have a statistically significant increase in fixations. 

\subsection{Augmenting Reading}
As a first step to explore reading comprehension more, it is evaluated if and how implicit text annotations using eye gaze can support second language learners and their teachers. 
Starting with giving readers quantified feedback about their behavior, answering simple quantitative questions: 
How fast do they read a paragraph? 
How much re-reading do they do? 
Yet, finally the aim is to give the reader feedback about their concentration and finally text comprehension level.

The current focus is set on paragraph based annotations, as these already can give valuable support to the learner and are feasible to implement with current technology. 
The initial set of annotations are inspired by lab internal discussions and by related work~\cite{Buscher:2008:GUG:1358628.1358805}.

In a prototype implementation, reading speed is highlighted by background color and intensity.
Slow speed with darker hue, faster speed with lighter hue. 
Reading speed is given by how long the participants eye gaze is in a paragraph region.

\begin{figure*}
    \includegraphics[width=\textwidth]{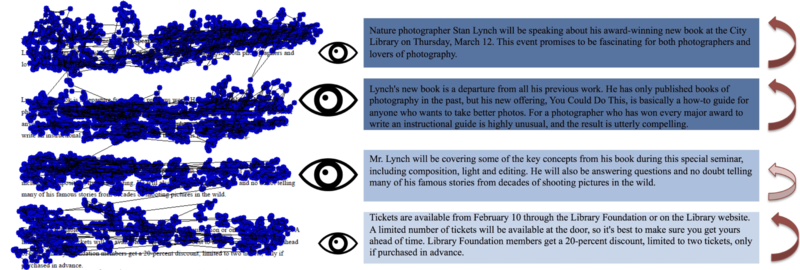}
    \includegraphics[width=\textwidth]{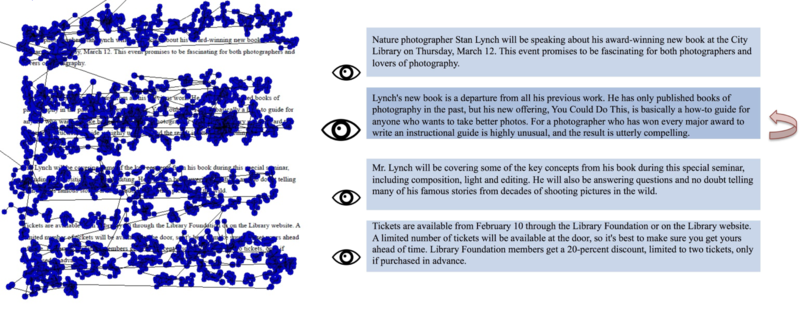}
   
  \caption{Eye-gaze annotated document for a participant with low English skills (first four paragraphs) and higher skills (second four paragraphs). 
  First we show the raw eyegaze as recorded by the eyetracker, then the annotated document. Shading shows the reading speed: the darker the slower. The arrows on the right show the amount of re-reading and the size of the eye next to the paragraph the number of fixation areas\cite{kunze2014}.}
  \label{fig:exp}
\end{figure*}

The amount of re-reading is estimated by comparing the line count of the paragraph with estimating the line count by using eye gaze using a method from Kunze et al.~\cite{kunze2013wordometer}.
The amount of re-reading is shown by an arrow pointing back up (cf. figure~\ref{fig:exp}).

Fixations are aggregated in larger fixation areas applying a filtering method from Busher et al.~\cite{Buscher:2008:GUG:1358628.1358805}. 
The number of fixation areas are shown as a eye icon next to the paragraph.

In Figure \ref{fig:exp}, we depict these annotations for a document read by students with good and poor English skills.
The good student performs less re-reading and has in general a fast reading speed. 
Although the differences between the two participants are easy to see, eye gaze is not only influenced by our expertise level, but also from fatigue and other mental states. 
Therefore, it's difficult to give comprehensive evaluations. 
Moving away from reading, we can also use cognitive activity recognition for implicitly tagging objects and events.
  
\subsection{Cognitive Tracking for the Masses}
A major problem of studies on cognitive activity recognition is that it is very difficult to make them representative, as sample sizes are relatively small (10 -20 participants). 
Problems also cover the activity recognition field in large and other information technology fields addressed. 
Dealing with cognitive tasks, this however is of additional weight. 
As seen from similar cognitive science and psychology studies, very large sample sizes are needed to assess the relations between tasks and cognitive activities, especially related with complex processes like learning.

One way to approach this problem is to provide affordable commodity devices to enable contributions from people towards the questions of intelligence amplification.

As eye trackers are still expensive and some people might not want to wear glasses, we should focus on alternative technologies that are already available or can be easily integrated into consumer devices, to enable cognitive task tracking.

\begin{figure}
    \centering
    \includegraphics[width=\columnwidth]{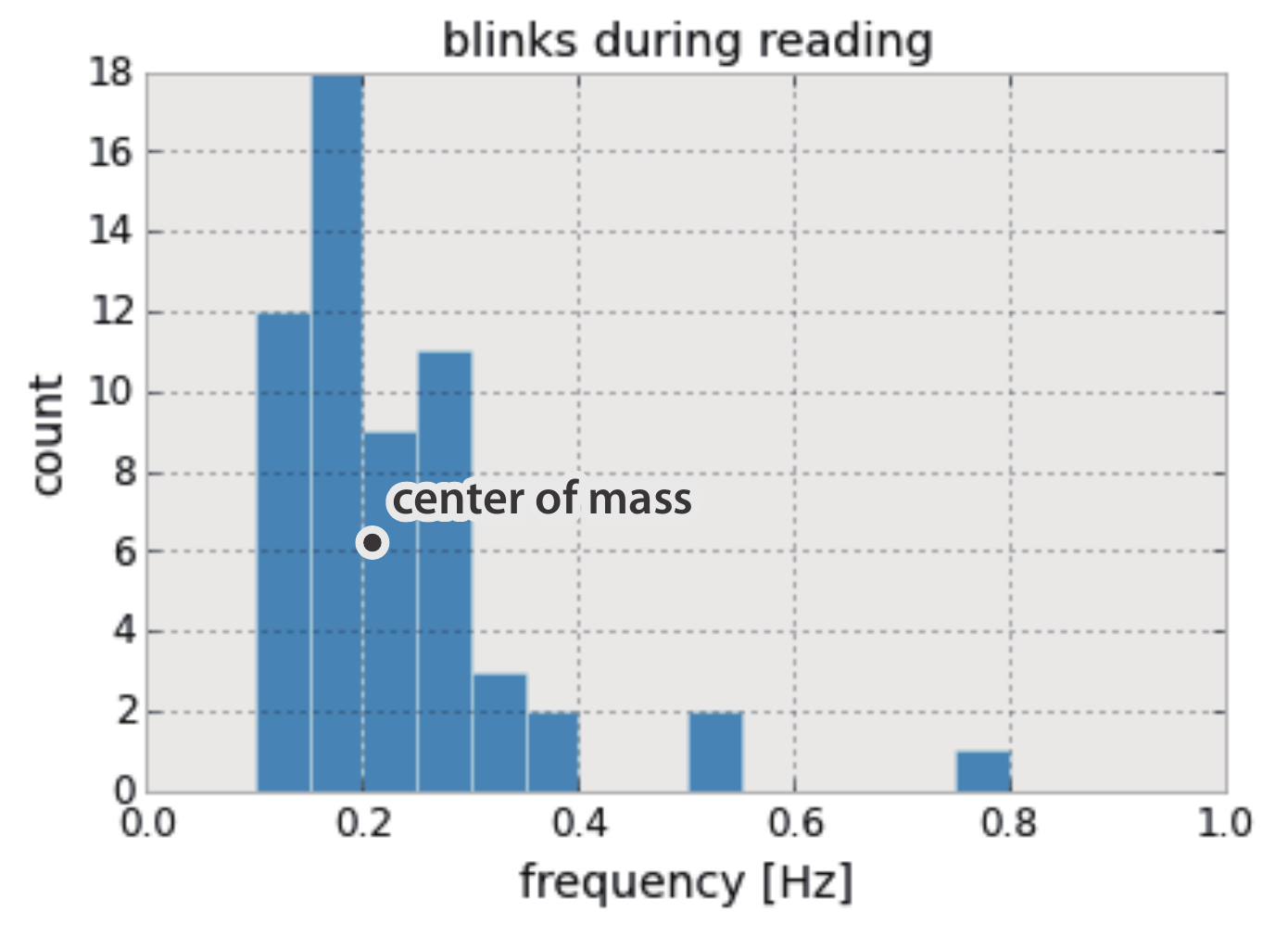}
    \includegraphics[width=\columnwidth]{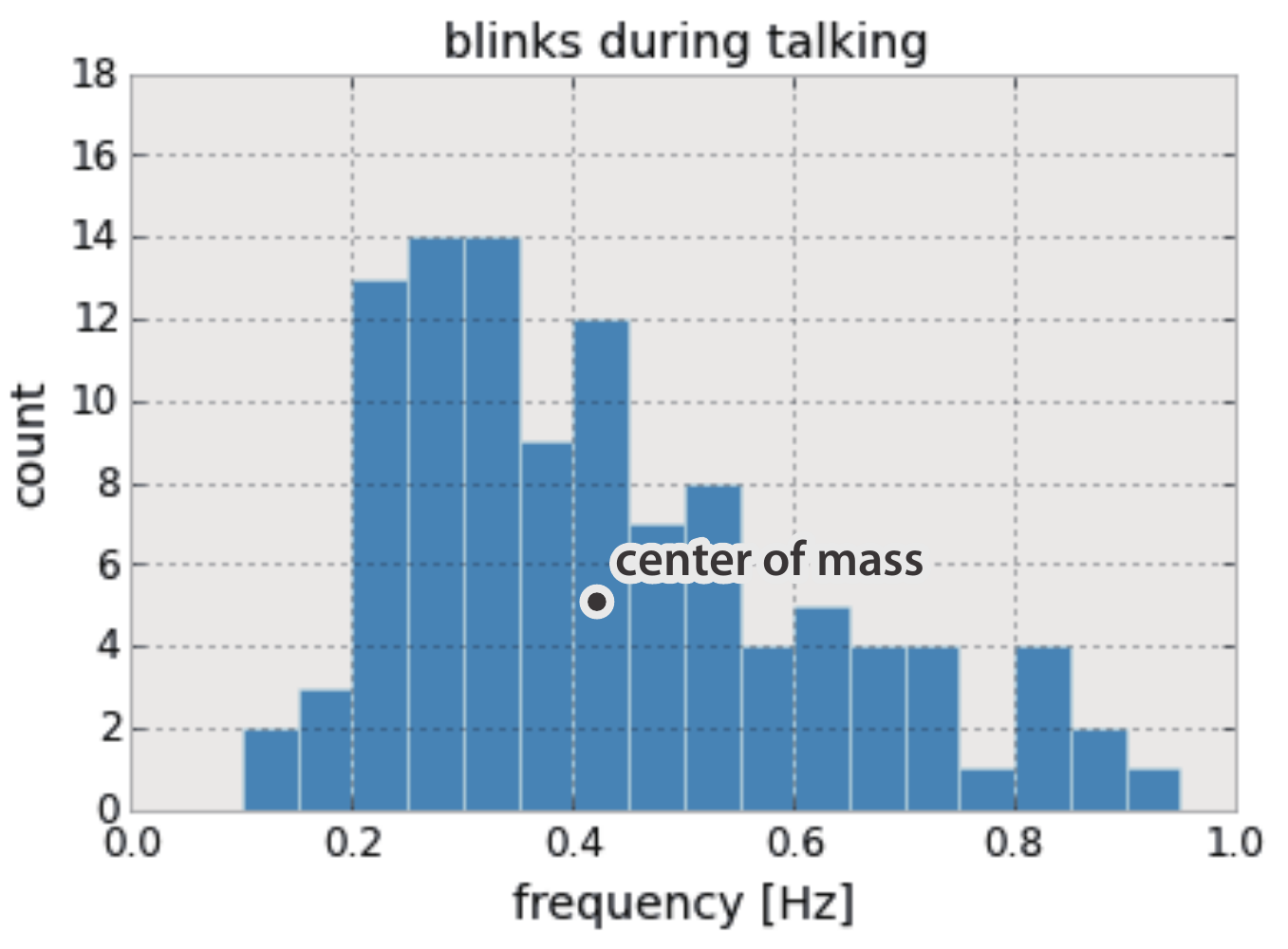}
   \caption{Blinking frequency recorded with Google Glass for reading (top) and talking to a person (bottom).}
   \label{blink}
\end{figure}

Additionally, head-mounted display computers, most prominently Google Glass, seem to get more and more commercial attention. 
They are a perfect tool for cognitive task analysis, as they are already worn on the head. 
A very simple sensor (infrared distance sensor from Google Glass) can for example measure eye blinks. 
Astonishingly, blinking frequency alone is already able to distinguish a couple of cognitive tasks (e.g. Reading versus Talking to a person, see Fig.~\ref{blink})~\cite{ishimaru2014blink}.

\section{Discussion and Future directions}\label{sectionFuture}
Activity recognition will increasingly focus on Parasitic and Sentiment Sensing paradigms. 
For device-free RF-based recognition, we expect that the diversity of sensors on devices can be greatly reduced as RF- and other environmental sources are capable to replace more specialised sensors with acceptable accuracy.
This will result in a simpler and thus cheaper design of consumer appliances with more accurate specialised sensing hardware reserved for professional devices.

Sentiment Sensing will receive considerable attention over the course of the next couple of years.
The knowledge on mental states will breed a number of new applications and challenges.

In addition, we expect that these sensing paradigms will increasingly be applied on non-expert off-the-shelf consumer hardware.
This development will foster a wide adaptation of these sensing paradigms and enable a number of novel applications as well as revenue for companies. 

\subsection{Environmental conditions}
Since parasitic sensing exploits environmental sensing sources, it is suggestive to monitor environmental conditions with such signals.
The sensing of traffic situations from environmental parasitic sources is gaining increased attention and might be fuelled also by vehicular communication, autonomous driving and pedestrian safety campaigns. 
But also other measures like, for instance, temperature can be sensed parasitically from RF. 

\subsubsection{Temperature}
As detailed in~\cite{Pervasive_Boano_2014}, the outside temperature impairs the capability of WiFi equipment, which might greatly reduce its transmission range.
By inversion of the same argument, the range of WiFi equipment will allow conclusions on the surrounding temperature.
While it is difficult to estimate the distance between a WiFi accesspoint (AP) and a wireless receiver directly, utilising changes in signal strength information from multiple APs should enable accurate prediction of environmental temperature.

\subsubsection{Sensing traffic situations}
Electromagnetic emission can be detected from a number of entities, including car engines. 
Regulation by EMC requires that emission from combustion engines fulfills strict requirements in the 30-1000 MHz range~\cite{RFSensing_Ruddle_2003}. 
But also for alternative power train road vehicles similar requirements apply~\cite{RFSensing_Ruddle_2003}.
These emissions are tested with standardized radiated emission tests such as CISPR~12~\cite{RFSensing_CISPR12_1997} or CISPR~22~\cite{RFSensing_CISPR22_1999}.

In~\cite{RFSensing_Dong_2006} it has been shown how RF emission from car's engines can be utilised in order to detect various car models.
The authors have been able to distinguish between three car models with an accuracy of 0.99 with the help of an Artificial Neural Network-based classifier.
For this, the ignition spark was the most characteristic event.
The characteristic features were identified over a frequency range of 2.5 GHz.

Kassem et al.~\cite{RFSensing_Kassem_2012} sense traffic situations by tracking frequency and speed of passing cars that intercept the direct line of sight between a pair of nodes on both sides of the road. 
Furthermore, Ding et al. demonstrated, how emissions from car engines can be utilised for passive traffic awareness utilising either roadside installations or also in-car modules~\cite{RFSensing_Ding_2011,RFSensing_Ding_2012}.
The authors have employed standard machine learning approaches in order to distinguish six traffic situations from roadside measurements and, in addition, the own-vehicule's speed with in-car measurements.
Recognition accuracy achieved in realistic environments were above 0.96 in all cases.
Possible further applications include the detection of traffic jams or also the number of cars waiting in front of a traffic light. 

\subsection{Sentiment and mental states}
As detailed above, sentiment and mental states are on the verge to being recognised from environmental and on-body sensors.
\subsubsection{Emotion}
Emotion can be inferred from body gesture and pose~\cite{Pervasive_Jaggarwal_2012} at least as accurately as from face~\cite{Emotion_Nguyen_2012,Emotion_Castellano_2008,Emotion_Meeren_2005}.
The role of human body in emotion expression has received support through evidences from psychology~\cite{Emotion_Walters_1986} and nonverbal communication \cite{Emotion_Dittmann_1978}. 
The importance of  bodily expression has also been confirmed for emotion detection \cite{Emotion_Wallbott_1998, Emotion_VanHeijnsbergen_2007,Emotion_Atkinson_2007}. 

Walter and Walk~\cite{Emotion_Walters_1986} revealed that emotion recognition from photos of postural expression, in which faces and hands were covered, was as accurate as recognition from facial expression alone.
Dynamic configurations of human body even hold greater amount of information as indicated by Bull in \cite{Emotion_Bull_1987}. 
He proved that body positions and motions could be recognized in terms of states including interest/boredom and agreement/disagreement. 
Some other studies went further by looking for the contribution of separated body parts to particular emotional states~\cite{Emotion_DeMeijer_1989,Emotion_Montepare_1999}.
Emotion can be recognized from non-trivial scenarios, such as simple daily-life actions \cite{Emotion_Crane_2007,Emotion_Bernhardt_2007} or recognition ability of infants \cite{Emotion_Lagerlof_2009}. 

It will be interesting to see how well RF information can be exploited in order to identify body gesture and pose and to classify this for human emotion classes.

\subsubsection{Attention}
Attention is an important measure in Computer-Human interaction. 
It determines for an interactive system the potential to impact the actions and decisions taken by an individual~\cite{AttentionMonitoring_Xu_2012}.
The same action of the same system might be considered either as annoyance or be appreciated as helpful depending on whether the individual was focusing part or all of her attention towards the system or not.
In the literature, we find various definitions that classify attention as well as its determining characteristics~\cite{AttentionMonitoring_Wu_2007, AttentionMonitoring_Wickens_1984}.
While the tracking of gaze is a commonly utilised measure of attention~\cite{AttentionMonitoring_Yonezawa_2007}, also other observable features may indicate attention. 
In general, aspects such as Saliency, Effort, Expectancy and Value are important indicators of attention~\cite{AttentionMonitoring_Wickens_2008, AttentionMonitoring_Wickens_1984,AttentionMonitoring_Xu_2012,AttentionMonitoring_Gollan_2011}.
This model was later extended to put a greater stress on the effort a person takes towards an object~\cite{AttentionMonitoring_Ferscha_2012}.
The authors also discuss various aspects of attention and identify as most distinguishing factors changes in walking speed, direction or orientation.

In~\cite{Pervasive_Shi_2014} it was investigated, how these properties, in particular location of a person, walking direction and walking speed or changes therein can be utilised for the monitoring and detection of attention. 
This was yet a preliminary study which lacked generalisation and high accuracy but we will see further improvements of attention recognition via RF soon. 

\subsection{Enhancing Recall and Focus}
Successfully tracking tasks, like emotion or attention
enables us to improve our cognitive abilities.
The ultimate goal of research conducted in these directions is to improve memory, concentration and finally decision making.

If we can track attention levels and cognitive load, we can identify the best times for the user to relax, learn, study or engage in spare-time activities, depending on their current cognitive state.

\subsection{Device-free RF-based recognition on consumer hardware}
\begin{figure}
    \centering
    \includegraphics[width=\columnwidth]{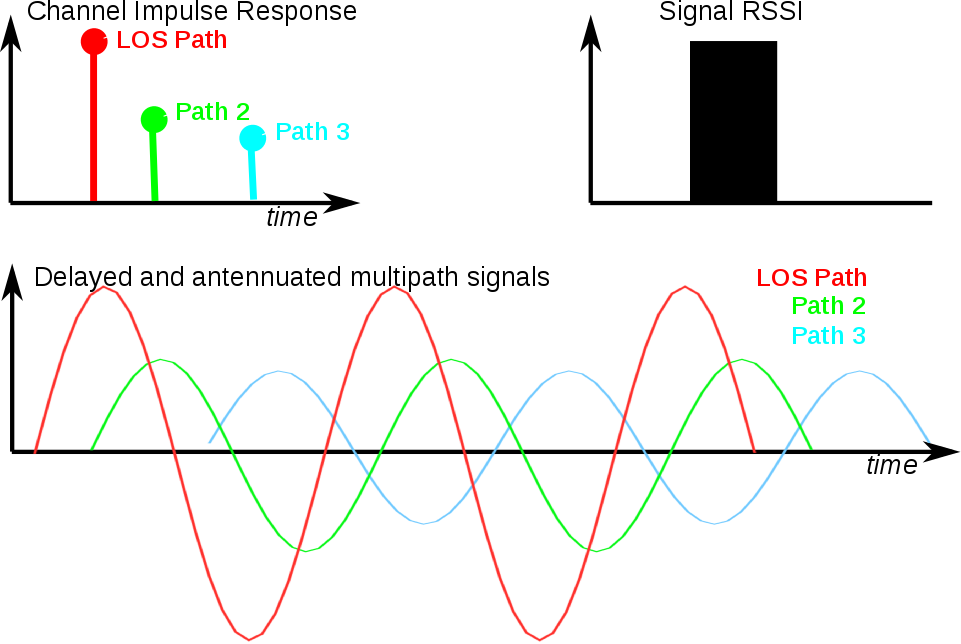}
   \caption{Multipath information contained in CSI PHY layer features in contrast to plain RSSI}
   \label{figureCSI}
\end{figure}

Currently, RF-based device-free recognition from continuous-signal based devices (such as e.g. SDR-nodes) can be considered as solved. 
Future directions are towards the recognition on consumer devices. 
With the introduction of OFDM to many wifi-class devices, some of the features, of SDR nodes, such as utilisation of multipath information can be incorporated from OFDM channel state information.
For WiFi-based indoor localisation, this has already been employed recently to achieve sub-meter accuracy~\cite{RFSensing_Wu_2012}.
In contrast to RSSI, the CSI contains channel response information as a PHY layer power feature~\cite{RFSensing_yang_2013}. 
Therefore, it becomes possible to discriminate multipath characteristics which hold the potential for more accurate classification of activities from RF. 
The utilisation of channel response was before recently only possible with sophisticated SDR hardware~\cite{RFSensing_Nerguizian_2006,RFSensing_Patwari_2007,RFSensing_Zhang_2008}. 
With introduction of Orthogonal Frequency Division Multiplexing (OFDM) for WiFi 802.11 a/g/n standards, this has, the channel response can now partially be extracted from off-the-shelf OFDM receivers, revealing amplitudes and phases of each subcarrier~\cite{RFSensing_Halperin_2011}. 
While RSSI is not able to capture the multipath effects in an environment, as depicted in figure~\ref{figureCSI}, the channel response available via CSI possesses finer grained frequency resolution and higher time resolution to distinguish multipath components.

Apart from this straightforward future research direction (which is already approached to-date by several groups), we can identify also more specific open research questions as follows.

\subsubsection{Empowering WiFi access points}
Authors have demonstrated the detection of several situations (for instance presence or crowd size) from RSSI information on a mobile phone~\cite{RFSensing_Sigg_2014}.
More interesting even is the estimation of crowd size or presence at a WiFi AP.
At the access point, the incoming packets originate from multiple devices at multiple locations.
In addition, traffic from an individual mobile device is typically much lower than the traffic generated by an AP.
It is not a-priori clear whether the snippets of RSSI-samples from distinct mobile devices are sufficient to estimate classes like crowd size or presence at a WiFi AP.
In particular, analysis of the evolution and fluctuation of the average RSSI level as well as normalisation of incoming flows regarding their signal strength might help to acquire such information.

\subsubsection{Activity recognition from 3G and 4G signals}
In~\cite{RFSensing_Sigg_2014} it was demonstrated how RSSI information from WiFi traffic can be utilised to identify environmental situations and gestures conducted in the proximity of a WiFi receiver.
Similarly, it will be possible to utilise 3G or 4G signals for the distinction of similar classes.
For this, however, the first step is the modification of the firmware for the 3G or 4G interface to allow access to signal strength information at higher frequency as this was done for the WiFi interface in~\cite{RFSensing_Sigg_2014}.

\section{Conclusion}
In this survey we have discussed recent advances in activity recognition which are leading towards two emerging sensing paradigms, namely Parasitic Sensing and Sentiment Sensing.
Both are fostered by the extreme increase in sensing devices in people's environments.
While classical sensing on mobile devices covers the surface of an individual's actions, namely her directly observable conditions, actions, movement and gestures, future sensing paradigms extend the reach of a device's perception.
Parasitic Sensing utilises noise of environmental, pre-installed systems and thereby captures stimuli from a device's near to mid-distance surroundings. 
On the other hand, sentiment Sensing reaches inwards, focusing on mental state and sentiment.
We see great potential for novel applications and revenue in both these paradigms.

Parasitic Sensing is fostered by the rise of the Internet of Things which will deploy a multitude of sensing and communicating devices in the environment.
In particular, these devices will feature an interface to the RF-channel which is why we envision this as the one universally employed sensor on such devices.
Apart from the already existing RF-noise to-day, IoT devices will generate significant additional traffic to transform the RF-interface into a rich sensing source for environmental activities. 

Sentiment Sensing in contrast benefits from a hype in novel body-worn devices, such as instrumented glasses or bio-sensors in a number of appliances.
Eyetracking is a rich source for the detection of a number of mental activities such as reading or also for the monitoring of attention or, for instance, fatigue.
Already today, products are announced which target this novel field of sensing\footnote{https://www.jins-jp.com/jinsmeme/en/}.
This will open new insights for applications and enable new fields of assistance for mobile devices.

We expect these sensing directions to flourish over the coming years and thereby to advance ubiquitous and pervasive sensing to new borders.


%

%


\ifCLASSOPTIONcaptionsoff
  \newpage
\fi



\bibliographystyle{IEEEtran}
%
%
%


\begin{thebibliography}{100}
\providecommand{\url}[1]{#1}
\csname url@samestyle\endcsname
\providecommand{\newblock}{\relax}
\providecommand{\bibinfo}[2]{#2}
\providecommand{\BIBentrySTDinterwordspacing}{\spaceskip=0pt\relax}
\providecommand{\BIBentryALTinterwordstretchfactor}{4}
\providecommand{\BIBentryALTinterwordspacing}{\spaceskip=\fontdimen2\font plus
\BIBentryALTinterwordstretchfactor\fontdimen3\font minus
  \fontdimen4\font\relax}
\providecommand{\BIBforeignlanguage}[2]{{%
\expandafter\ifx\csname l@#1\endcsname\relax
\typeout{** WARNING: IEEEtran.bst: No hyphenation pattern has been}%
\typeout{** loaded for the language `#1'. Using the pattern for}%
\typeout{** the default language instead.}%
\else
\language=\csname l@#1\endcsname
\fi
#2}}
\providecommand{\BIBdecl}{\relax}
\BIBdecl

\bibitem{Pervasive_Ferscha_2014}
A.~Ferscha, ``Attention, please!'' \emph{Pervasive Computing, IEEE}, vol.~13,
  no.~1, pp. 48--54, 2014.

\bibitem{RFsensing_Pu_2013}
Q.~Pu, S.~Gupta, S.~Gollakota, and S.~Patel, ``Whole-home gesture recognition
  using wireless signals,'' in \emph{The 19th Annual International Conference
  on Mobile Computing and Networking (Mobicom'13)}, 2013.

\bibitem{Pervasive_Adib_2013}
F.~Adib and D.~Katabi, ``See through walls with wi-fi,'' in \emph{ACM
  SIGCOMM'13}, 2013.

\bibitem{Cryptography_Schuerman_2011}
D.~Schuermann and S.~Sigg, ``Secure communication based on ambient audio,''
  \emph{IEEE Transactions on mobile computing}, vol.~12, no.~2, 2013.

\bibitem{Cryptography_Madiseh_2008}
M.~G. Madiseh, M.~L. McGuire, S.~S. Neville, L.~Cai, and M.~Horie, ``Secret key
  generation and agreement in uwb communication channels,'' in
  \emph{Proceedings of the 51st International Global Communications Conference
  (Globecom)}, 2008.

\bibitem{Pervasive_Ishimaru_2014}
S.~Ishimaru, K.~Kunze, K.~Kise, J.~Weppner, A.~Dengel, P.~Lukowicz, and
  A.~Bulling, ``In the blink of an eye: combining head motion and eye blink
  frequency for activity recognition with google glass,'' in \emph{Proceedings
  of the 5th Augmented Human International Conference}.\hskip 1em plus 0.5em
  minus 0.4em\relax ACM, 2014, p.~15.

\bibitem{Pervasive_Kunze_2013}
K.~Kunze, Y.~Utsumi, Y.~Shiga, K.~Kise, and A.~Bulling, ``I know what you are
  reading: recognition of document types using mobile eye tracking,'' in
  \emph{Proceedings of the 17th annual international symposium on International
  symposium on wearable computers}.\hskip 1em plus 0.5em minus 0.4em\relax ACM,
  2013, pp. 113--116.

\bibitem{Pervasive_Castellano_2008}
G.~Castellano, L.~Kessous, and G.~Caridakis, ``Emotion recognition through
  multiple modalities: face, body gesture, speech,'' in \emph{Affect and
  emotion in human-computer interaction}.\hskip 1em plus 0.5em minus
  0.4em\relax Springer, 2008, pp. 92--103.

\bibitem{Pervasive_Jaggarwal_2012}
A.~Jaggarwal and R.~L. Canosa, ``Emotion recognition using body gesture and
  pose,'' Tech. Rep., 2012, available online:
  http://www.cs.rit.edu/$\sim$axj4159/papers\_march/report\_1.pdf.

\bibitem{Pervasive_Roggen_2009}
D.~Roggen, K.~Forster, A.~Calatroni, T.~Holleczek, Y.~Fang, G.~Troster,
  P.~Lukowicz, G.~Pirkl, D.~Bannach, K.~Kunze, A.~Ferscha, C.~Holzmann,
  A.~Riener, R.~Chavarriaga, and J.~del R~Millan, ``Opportunity: Towards
  opportunistic activity and context recognition systems,'' in \emph{World of
  Wireless, Mobile and Multimedia Networks Workshops, 2009. WoWMoM 2009. IEEE
  International Symposium on a}, 2009, pp. 1--6.

\bibitem{Pervasive_Cheng_2013}
J.~Cheng, B.~Zhou, K.~Kunze, C.~C. Rheinl\"{a}nder, S.~Wille, N.~Wehn,
  J.~Weppner, and P.~Lukowicz, ``Activity recognition and nutrition monitoring
  in every day situations with a textile capacitive neckband,'' in
  \emph{Proceedings of the 2013 ACM Conference on Pervasive and Ubiquitous
  Computing Adjunct Publication}, ser. UbiComp '13 Adjunct, 2013, pp. 155--158.

\bibitem{Pervasive_Chen_2012}
L.~Chen, C.~Nugent, and H.~Wang, ``A knowledge-driven approach to activity
  recognition in smart homes,'' \emph{Knowledge and Data Engineering, IEEE
  Transactions on}, vol.~24, no.~6, pp. 961--974, 2012.

\bibitem{Pervasive_Chen_2012-2}
L.~Chen, J.~Hoey, C.~D. Nugent, D.~J. Cook, and Z.~Yu, ``Sensor-based activity
  recognition,'' \emph{IEEE Transaction on Systems, Man, and Cybernetics, Part
  C: Applications and Reviews}, vol.~PP, no.~99, pp. 1 --19, 2012.

\bibitem{Opportunistic_Campbell_2008}
A.~T. Campbell, S.~B. Eisenman, N.~D. Lane, E.~Miluzzo, R.~A. Peterson, H.~Lu,
  X.~Zheng, M.~Musolesi, K.~Fodor, and G.-S. Ahn, ``The rise of people-centric
  sensing,'' \emph{Internet Computing, IEEE}, vol.~12, no.~4, pp. 12--21, 2008.

\bibitem{OpportunisticSensing_Lane_2008}
N.~D. Lane, S.~B. Eisenman, M.~Musolesi, E.~Miluzzo, and A.~T. Campbell,
  ``Urban sensing systems: opportunistic or participatory?'' in
  \emph{Proceedings of the 9th workshop on Mobile computing systems and
  applications}.\hskip 1em plus 0.5em minus 0.4em\relax ACM, 2008, pp. 11--16.

\bibitem{OpportunisticSensing_Kapadia_2008}
A.~Kapadia, N.~Triandopoulos, C.~Cornelius, D.~Peebles, and D.~Kotz,
  ``Anonysense: Opportunistic and privacy-preserving context collection,'' in
  \emph{Pervasive Computing}.\hskip 1em plus 0.5em minus 0.4em\relax Springer,
  2008, pp. 280--297.

\bibitem{OpportunisticSensing_Roggen_2009}
D.~Roggen, K.~Forster, A.~Calatroni, T.~Holleczek, Y.~Fang, G.~Troster,
  P.~Lukowicz, G.~Pirkl, D.~Bannach, K.~Kunze \emph{et~al.}, ``Opportunity:
  Towards opportunistic activity and context recognition systems,'' in
  \emph{World of Wireless, Mobile and Multimedia Networks \& Workshops, 2009.
  WoWMoM 2009. IEEE International Symposium on a}.\hskip 1em plus 0.5em minus
  0.4em\relax IEEE, 2009, pp. 1--6.

\bibitem{OpportunisticSensing_Kurz_2011}
M.~Kurz, G.~H{\"o}lzl, A.~Ferscha, A.~Calatroni, D.~Roggen, G.~Tr{\"o}ster,
  H.~Sagha, R.~Chavarriaga, J.~d.~R. Mill{\'a}n, D.~Bannach \emph{et~al.},
  ``The opportunity framework and data processing ecosystem for opportunistic
  activity and context recognition,'' \emph{International Journal of Sensors,
  Wireless Communications and Control, Special Issue on Autonomic and
  Opportunistic Communications}, vol.~1, 2011.

\bibitem{OpportunisticSensing_Shin_2011}
M.~Shin, C.~Cornelius, D.~Peebles, A.~Kapadia, D.~Kotz, and N.~Triandopoulos,
  ``Anonysense: A system for anonymous opportunistic sensing,'' \emph{Pervasive
  and Mobile Computing}, vol.~7, no.~1, pp. 16--30, 2011.

\bibitem{ParticipatorySensing_Burke_2006}
J.~A. Burke, D.~Estrin, M.~Hansen, A.~Parker, N.~Ramanathan, S.~Reddy, and
  M.~B. Srivastava, ``Participatory sensing,'' 2006.

\bibitem{Pervasive_patel_2008}
S.~N. Patel, M.~S. Reynolds, and G.~D. Abowd, ``Detecting human movement by
  differential air pressure sensing in hvac system ductwork: An exploration in
  infrastructure mediated sensing,'' in \emph{Proceedings of the 6th
  International Conference on Pervasive Computing (Pervasive 2008)}, 2008.

\bibitem{Pervasive_patel_2008b}
S.~N. Patel, \emph{Infrastructure mediated sensing}.\hskip 1em plus 0.5em minus
  0.4em\relax ProQuest, 2008.

\bibitem{ContextAwareness_Kunze_2007}
K.~Kunze and P.~Lukowicz, ``Symbolic object localization through active
  sampling of acceleration and sound signatures,'' in \emph{Proceedings of the
  9th International Conference on Ubiquitous Computing}, 2007.

\bibitem{RFSensing_Sigg_2014}
S.~Sigg, U.~Blanke, and G.~Troester, ``The telepathic phone: Frictionless
  activity recognition from wifi-rssi,'' in \emph{IEEE International Conference
  on Pervasive Computing and Communications (PerCom)}, ser. PerCom '14, 2014.

\bibitem{Antifakos:2002p8030}
\BIBentryALTinterwordspacing
S.~Antifakos, F.~Michahelles, and B.~Schiele, ``Proactive instructions for
  furniture assembly,'' \emph{LECTURE NOTES IN COMPUTER SCIENCE}, Jan 2002.
  [Online]. Available:
  \url{http://www.springerlink.com/index/5B4PAXAQHCFM96F9.pdf}
\BIBentrySTDinterwordspacing

\bibitem{heinz2006uws}
E.~A. Heinz, K.~Kunze, M.~Gruber, D.~Bannach, and P.~Lukowicz, ``{Using
  Wearable Sensors for Real-time Recognition Tasks in Games of Martial Arts--An
  Initial Experiment},'' in \emph{Proceedings of the 2nd IEEE Symposium on
  Computational Intelligence and Games (CIG 2006), {\sf Reno/Lake Tahoe, NV}},
  2006, pp. 98--102.

\bibitem{kunze2008dsd}
K.~Kunze and P.~Lukowicz, ``{Dealing with sensor displacement in motion-based
  onbody activity recognition systems},'' in \emph{Proceedings of the 10th
  international conference on Ubiquitous computing}.\hskip 1em plus 0.5em minus
  0.4em\relax ACM New York, NY, USA, 2008, pp. 20--29.

\bibitem{kunze2013activity}
K.~Kunze, M.~Iwamura, K.~Kise, S.~Uchida, and S.~Omachi, ``Activity recognition
  for the mind: Toward a cognitive ``quantified self'','' \emph{Computer},
  vol.~46, no.~10, pp. 0105--108, 2013.

\bibitem{matthews2007wearable}
R.~Matthews, N.~J. McDonald, P.~Hervieux, P.~J. Turner, and M.~A. Steindorf,
  ``A wearable physiological sensor suite for unobtrusive monitoring of
  physiological and cognitive state,'' in \emph{Engineering in Medicine and
  Biology Society, 2007. EMBS 2007. 29th Annual International Conference of the
  IEEE}.\hskip 1em plus 0.5em minus 0.4em\relax IEEE, 2007, pp. 5276--5281.

\bibitem{Pervasive_bulling_2009}
A.~Bulling, D.~Roggen, and G.~Tr{\"o}ster, \emph{Wearable EOG goggles:
  eye-based interaction in everyday environments}.\hskip 1em plus 0.5em minus
  0.4em\relax ACM, 2009.

\bibitem{Pervasive_Chang_2009}
K.-h. Chang, J.~Hightower, and B.~Kveton, ``Inferring identity using
  accelerometers in television remote controls,'' in \emph{Pervasive
  Computing}, ser. Lecture Notes in Computer Science, H.~Tokuda, M.~Beigl,
  A.~Friday, A.~Brush, and Y.~Tobe, Eds., vol. 5538.\hskip 1em plus 0.5em minus
  0.4em\relax Springer Berlin Heidelberg, 2009, pp. 151--167.

\bibitem{Pervasive_Laerhoven_2004}
K.~Van~Laerhoven and H.-W. Gellersen, ``Spine versus porcupine: a study in
  distributed wearable activity recognition,'' in \emph{Wearable Computers,
  2004. ISWC 2004. Eighth International Symposium on}, vol.~1, 2004, pp. 142 --
  149.

\bibitem{Pervasive_Miluzzo_2008}
E.~Miluzzo, N.~D. Lane, K.~Fodor, R.~Peterson, H.~Lu, M.~Musolesi, S.~B.
  Eisenman, X.~Zheng, and A.~T. Campbell, ``Sensing meets mobile social
  networks: The design, implementation and evaluation of the cenceme
  application,'' in \emph{Proceedings of the 6th ACM Conference on Embedded
  Network Sensor Systems}, ser. SenSys '08, 2008, pp. 337--350.

\bibitem{Pervasive_Laerhoven_2008}
K.~Van~Laerhoven, D.~Kilian, and B.~Schiele, ``Using rhythm awareness in
  long-term activity recognition,'' in \emph{Proceedings of the 2008 12th IEEE
  International Symposium on Wearable Computers}, ser. ISWC '08, 2008, pp.
  63--66.

\bibitem{Pervasive_Welbourne_2005}
E.~Welbourne, J.~Lester, A.~LaMarca, and G.~Borriello, ``Mobile context
  inference using low-cost sensors,'' in \emph{Location- and
  Context-Awareness}, ser. Lecture Notes in Computer Science, vol. 3479.\hskip
  1em plus 0.5em minus 0.4em\relax Springer Berlin Heidelberg, pp. 254--263.

\bibitem{5952}
L.~Bao and S.~S. Intille, ``Activity recognition from user-annotated
  acceleration data,'' in \emph{Proceedings of the 2nd International conference
  on Pervasive Computing}, April 2004.

\bibitem{Pervasive_Lester_2006}
J.~Lester, T.~Choudhury, and G.~Borriello, ``A practical approach to
  recognizing physical activities,'' in \emph{Pervasive Computing}, ser.
  Lecture Notes in Computer Science, K.~Fishkin, B.~Schiele, P.~Nixon, and
  A.~Quigley, Eds.\hskip 1em plus 0.5em minus 0.4em\relax Springer Berlin /
  Heidelberg, 2006, vol. 3968, pp. 1--16.

\bibitem{Pervasive_Tapia_2007}
E.~Tapia, S.~Intille, W.~Haskell, K.~Larson, J.~Wright, A.~King, and
  R.~Friedman, ``Real-time recognition of physical activities and their
  intensities using wireless accelerometers and a heart rate monitor,'' in
  \emph{Wearable Computers, 2007 11th IEEE International Symposium on}, 2007,
  pp. 37--40.

\bibitem{Pervasive_Gordon_2010b}
D.~Gordon, H.~Schmidtke, M.~Beigl, and G.~von Zengen, ``A novel micro-vibration
  sensor for activity recognition: Potential and limitations,'' in
  \emph{Wearable Computers (ISWC), 2010 International Symposium on}, 2010, pp.
  1--8.

\bibitem{Pervasive_Pirkl_2008}
G.~Pirkl, K.~Stockinger, K.~Kunze, and P.~Lukowicz, ``Adapting magnetic
  resonant coupling based relative positioning technology for wearable
  activitiy recogniton,'' in \emph{Wearable Computers, 2008. ISWC 2008. 12th
  IEEE International Symposium on}, 2008, pp. 47--54.

\bibitem{Pervasive_Yannakakis_2008}
G.~Yannakakis, J.~Hallam, and H.~Lund, ``{Entertainment capture through heart
  rate activity in physical interactive playgrounds},'' \emph{User Modeling and
  User-Adapted Interaction}, vol.~18, pp. 207--243, 2008.

\bibitem{Pervasive_Intille_2003}
S.~Intille, E.~Tapia, J.~Rondoni, J.~Beaudin, C.~Kukla, S.~Agarwal, L.~Bao, and
  K.~Larson, ``{Tools for Studying Behavior and Technology in Natural
  Settings},'' in \emph{{UbiComp 2003: Ubiquitous Computing}}, ser. {Lecture
  Notes in Computer Science}.\hskip 1em plus 0.5em minus 0.4em\relax Springer
  Berlin Heidelberg, 2003, vol. 2864, pp. 157--174.

\bibitem{Pervasive_Michahelles_2003}
F.~Michahelles, P.~Matter, A.~Schmidt, and B.~Schiele, ``{Applying wearable
  sensors to avalanche rescue},'' \emph{Computers \& Graphics}, vol.~27, no.~6,
  pp. 839--847, 2003.

\bibitem{Pervasive_Christopoulou_2005}
E.~Christopoulou, C.~Goumopoulos, and A.~Kameas, ``An ontology-based context
  management and reasoning process for ubicomp applications,'' in
  \emph{Proceedings of the 2005 Joint Conference on Smart Objects and Ambient
  Intelligence: Innovative Context-aware Services: Usages and Technologies},
  ser. sOc-EUSAI '05, 2005, pp. 265--270.

\bibitem{Pervasive_seo_2004}
J.~W. Seo and K.~Park, ``The development of a ubiquitous health house in south
  korea,'' in \emph{UbiComp}, 2004.

\bibitem{Pervasive_lo_2005}
B.~Lo, S.~Thiemjarus, R.~King, and G.-Z. Yang, ``Body sensor network-a wireless
  sensor platform for pervasive healthcare monitoring,'' in \emph{The 3rd
  International Conference on Pervasive Computing}, vol.~13, 2005, pp. 77--80.

\bibitem{Pervasive_gamecho_2013}
B.~Gamecho, L.~Gardeazabal, and J.~Abascal, ``Combination and abstraction of
  sensors for mobile context-awareness,'' in \emph{Proceedings of the 2013 ACM
  conference on Pervasive and ubiquitous computing adjunct publication}.\hskip
  1em plus 0.5em minus 0.4em\relax ACM, 2013, pp. 1417--1422.

\bibitem{Pervasive_Amft_2009}
O.~Amft, M.~Kusserow, and G.~Troster, ``Bite weight prediction from acoustic
  recognition of chewing,'' \emph{Biomedical Engineering, IEEE Transactions
  on}, vol.~56, no.~6, pp. 1663--1672, 2009.

\bibitem{Pervasive_Ogris_2007}
G.~Ogris, M.~Kreil, and P.~Lukowicz, ``Using fsr based muscule activity
  monitoring to recognize manipulative arm gestures,'' in \emph{Wearable
  Computers, 2007 11th IEEE International Symposium on}, 2007, pp. 45--48.

\bibitem{Pervasive_Bahl_2000}
P.~Bahl and V.~Padmanabhan, ``Radar: an in-building rf-based user location and
  tracking system,'' in \emph{Proceedings of the 19th IEEE International
  Conference on Computer Communications (Infocom)}, 2000.

\bibitem{Pervasive_Otsason_2005}
V.~Otsason, A.~Varshavsky, A.~LaMarca, and E.~de~Lara, ``Accurate gsm indoor
  localisation,'' in \emph{Proceedings of the 7th ACM International Conference
  on Ubiquitous Computing (Ubicomp 2005)}, 2005.

\bibitem{Pervasive_Varshavsky_2007}
A.~Varshavsky, E.~de~Lara, J.~Hightower, A.~LaMarca, and V.~Otsason, ``Gsm
  indoor localization,'' \emph{Pervasive and Mobile Computing}, vol.~3, 2007.

\bibitem{Pervasive_Krumm_2003}
J.~Krumm and G.~Cermak, ``Rightspot: A novel sense of location for a smart
  personal object,'' in \emph{Proceedings of the 5th ACM International
  Conference on Ubiquitous Computing (Ubicomp 2003)}, 2003.

\bibitem{Pervasive_Youssef_2005}
A.~Youssef, J.~Krumm, E.~Miller, G.~Cermak, and E.~Horvitz, ``Computing
  location from ambient fm radio signals,'' in \emph{Proceedings of the IEEE
  Wireless Communications and Networking Conference}, 2005.

\bibitem{Pervasive_Chaquet_2013}
J.~M. Chaquet, E.~J. Carmona, and A.~Fern\'{a}Ndez-Caballero, ``A survey of
  video datasets for human action and activity recognition,'' \emph{Comput.
  Vis. Image Underst.}, vol. 117, no.~6, pp. 633--659, Jun. 2013.

\bibitem{ActivityRecognition_Aggarwal_2011}
J.~Aggarwal and M.~Ryoo, ``Human activity analysis: A review,'' \emph{ACM
  Computing Surveys}, vol.~43, no.~3, pp. 16:1--16:43, Apr. 2011.

\bibitem{LocationTracking_Cai_1998}
Q.~Cai and J.~K. Aggarwal, ``Automatic tracking of human motion in indoor
  scenes across multiple synchronized video streams,'' in \emph{Proceedings of
  the Sixth International Conference on Computer Vision}, 1998.

\bibitem{Pervasive_Han_2005}
J.~Han and B.~Bhanu, ``Human activity recognition in thermal infrared
  imagery,'' in \emph{Computer Vision and Pattern Recognition - Workshops,
  2005. CVPR Workshops. IEEE Computer Society Conference on}, 2005, pp. 17--17.

\bibitem{5836}
H.-W. Gellersen, G.~Kortuem, A.~Schmidt, and M.~Beigl, ``Physical prototyping
  with smart-its,'' \emph{IEEE Pervasive computing}, vol.~4, no. 1536-1268, pp.
  10--18, 2004.

\bibitem{Pervasive_Enokibori_2013}
Y.~Enokibori, A.~Suzuki, H.~Mizuno, Y.~Shimakami, and K.~Mase, ``E-textile
  pressure sensor based on conductive fiber and its structure,'' in
  \emph{Proceedings of the 2013 ACM Conference on Pervasive and Ubiquitous
  Computing Adjunct Publication}, ser. UbiComp '13 Adjunct, 2013, pp. 207--210.

\bibitem{5840}
M.~Beigl, A.~Krohn, T.~Zimmer, and C.~Decker, ``Typical sensors needed in
  ubiquitous and pervasive computing,'' in \emph{First International Workshop
  on Networked Sensing Systems}, ser. Society of Instrument and Control
  Engineers, vol.~4, June 2004, pp. 153--158.

\bibitem{LocationTracking_Robert_2000}
J.~O. Robert and D.~A. Gregory, ``The smart floor: a mechanism for natural user
  identification and tracking,'' in \emph{Proceedings of the CHI 2000
  Conference on Human Factors in Computing Systems}, 2000.

\bibitem{5077}
R.~Want, A.~Hopper, V.~Falcao, and J.~Gibbons, ``The active badge location
  system,'' in \emph{ACM Transactions on Information Systems}, vol.~1, no.~10,
  1992, pp. 91--102.

\bibitem{LocationTracking_Priyantha_2000}
N.~B. Priyantha, A.~Chakraborty, and H.~Balakrishnan, ``The cricket
  location-support system,'' in \emph{Proceedings of the Sixth Annual
  International Conference on Mobile Computing and Networking}, 2000.

\bibitem{Pervasive_Ren_2011}
Z.~Ren, J.~Meng, J.~Yuan, and Z.~Zhang, ``Robust hand gesture recognition with
  kinect sensor,'' in \emph{Proceedings of the 19th ACM International
  Conference on Multimedia}, 2011, pp. 759--760.

\bibitem{Pervasive_Panger_2012}
G.~Panger, ``Kinect in the kitchen: Testing depth camera interactions in
  practical home environments,'' in \emph{CHI '12 Extended Abstracts on Human
  Factors in Computing Systems}, ser. CHI EA '12, 2012, pp. 1985--1990.

\bibitem{Pervasive_Motta_2013}
T.~Motta and L.~Nedel, ``Deviceless gestural interaction for public displays,''
  in \emph{Proceedings of the 2013 XV Symposium on Virtual and Augmented
  Reality}, ser. SVR '13, 2013, pp. 175--184.

\bibitem{Pervasive_Patel_2007}
S.~N. Patel, T.~Robertson, J.~A. Kientz, M.~S. Reynolds, and G.~D. Abowd, ``At
  the flick of a switch: Detecting and classifying unique electrical events on
  the residential power line,'' in \emph{Proceedings of the 9th International
  Conference on Ubiquitous Computing (UbiComp 2007)}, 2007, pp. 271--288.

\bibitem{RFSensing_Gupta_2010}
S.~Gupta, M.~S. Reynolds, and S.~N. Patel, ``Electrisense: Single-point sensing
  using emi for electrical event detection and classificaiton in the home,'' in
  \emph{Proceedings of the 13th international conference on Ubiquitous
  computing}, 2010.

\bibitem{Pervasive_Gupta_2011}
S.~Gupta, K.-Y. Chen, M.~S. Reynolds, and S.~N. Patel, ``Lightwave: Using
  compact fluorescent lights as sensors,'' in \emph{Proceedings of the 13th
  international conference on Ubiquitous computing}, 2011.

\bibitem{Pervasive_Campbell_2010}
A.~T. Campbell, E.~Larson, G.~Cohn, J.~Froehlich, R.~Alcaide, and S.~N. Patel,
  ``Wattr: A method for self-prowered wireless sensing of water activity in the
  home,'' in \emph{Proceedings of the 12th international conference on
  Ubiquitous computing}, 2010.

\bibitem{Pervasive_Thomaz_2012}
E.~Thomaz, V.~Bettadapura, G.~Reyes, M.~Sandesh, G.~Schindler, T.~Ploetz, G.~D.
  Abowd, and I.~Essa, ``Recognizing water-based activities in the home through
  infrastructure-mediated sensing,'' in \emph{Proceedings of the 14th ACM
  International Conference on Ubiquitous Computing (Ubicomp 2012)}, 2012.

\bibitem{Pervasive_Froehlich_2009}
J.~E. Froehlich, E.~Larson, T.~Campbell, C.~Haggerty, J.~Fogarty, and S.~N.
  Patel, ``Hydrosense: Infrastructure-mediated single-point sensing of
  whole-home water activity,'' in \emph{Proceedings of the 11th International
  Conference on Ubiquitous Computing}, ser. Ubicomp '09, 2009, pp. 235--244.

\bibitem{Pervasive_Cohn_2010b}
G.~Cohn, S.~Gupta, J.~Froehlich, E.~Larson, and S.~Patel, ``{GasSense:
  Appliance-Level, Single-Point Sensing of Gas Activity in the Home},'' in
  \emph{{Pervasive Computing}}, ser. {Lecture Notes in Computer Science},
  P.~Flor{\'e}en, A.~Kr{\"u}ger, and M.~Spasojevic, Eds.\hskip 1em plus 0.5em
  minus 0.4em\relax Springer Berlin Heidelberg, vol. 6030, pp. 265--282.

\bibitem{Pervasive_Cohn_2011}
G.~Cohn, D.~Morris, S.~N. Patel, and D.~S. Tan, ``Your noise is my command:
  Sensing gestures using the body as an antenna,'' in \emph{Proceedings of the
  SIGCHI Conference on Human Factors in Computing Systems}, ser. CHI '11, 2011,
  pp. 791--800.

\bibitem{Pervasive_Cohn_2012}
------, ``Humantenna: Using the body as an antenna for real-time whole-body
  interaction,'' in \emph{Proceedings of ACM CHI 2012}, 2012.

\bibitem{DeviceFreeRecognition_Tan_2005}
D.~Tan, H.~Sun, Y.~Lu, M.~Lesturgie, and H.~Chan, ``Passive radar using global
  system for mobile communication signal: theory, implementation and
  measurements,'' \emph{IEE Proceedings - Radar, Sonar and Navigation}, vol.
  152, no.~3, pp. 116--123, 2005.

\bibitem{DeviceFreeRecognition_Colone_2012}
F.~Colone, P.~Falcone, C.~Bongioanni, and P.~Lombardo, ``Wifi-based passive
  bistatic radar: Data processing schemes and experimental results,''
  \emph{IEEE Transactions on Aerospace and Electronic Systems}, vol.~48, no.~2,
  pp. 1061--1079, 2012.

\bibitem{Pervasive_Youssef_2007}
M.~Youssef, M.~Mah, and A.~Agrawala, ``Challenges: Device-free passive
  localisation for wireless environments,'' in \emph{Proceedings of the 13th
  annual ACM international Conference on Mobile Computing and Networking
  (MobiCom 2007)}, 2007, pp. 222--229.

\bibitem{RFSensing_Zhang_2011}
D.~Zhang, Y.~Liu, and L.~Ni, ``Rass: A real-time, accurate and scalable system
  for tracking transceiver-free objects,'' in \emph{Proceedings of the 9th IEEE
  International Conference on Pervasive Computing and Communications
  (PerCom2011)}, 2011.

\bibitem{RFSensing_Ding_2011}
Y.~Ding, B.~Banitalebi, T.~Miyaki, and M.~Beigl, ``Rftraffic: Passive traffic
  awareness based on emitted rf noise from the vehicles,'' in \emph{ITS
  Telecommunications (ITST), 2011 11th International Conference on}, aug. 2011,
  pp. 393 --398.

\bibitem{Pervasive_Shi_2014}
S.~Shi, S.~Sigg, W.~Zhao, and Y.~Ji, ``Monitoring of attention from ambient
  fm-radio signals,'' \emph{IEEE Pervasive Computing, Special Issue - Managing
  Attention in Pervasive Environments}, Jan-March 2014.

\bibitem{RFSensing_Patwari_2011b}
\BIBentryALTinterwordspacing
N.~Patwari and J.~Wilson, ``Rf sensor networks for device-free localization:
  Measurements, models, and algorithms,'' \emph{Proceedings of the IEEE},
  vol.~98, no.~11, pp. 1961--1973, 2010. [Online]. Available:
  \url{http://ieeexplore.ieee.org/xpls/ abs\_all.jsp?arnumber=5523907}
\BIBentrySTDinterwordspacing

\bibitem{Pervasive_Berchtold_2010}
M.~Berchtold, M.~Budde, D.~Gordon, H.~R. Schmidtke, and M.~Beigl, ``Actiserv:
  Activity recognition service for mobile phones,'' in \emph{International
  Symposium on Wearable Computers (ISWC)}, 2010, pp. 1--8.

\bibitem{Pervasive_Bao_2004}
L.~Bao and S.~S. Intille, ``Activity recognition from user-annotated
  acceleration data,'' in \emph{Proceedings of PERVASIVE 2004}, vol. LNCS 3001,
  2004.

\bibitem{Pervasive_Schmidt_2012}
A.~Schmidt, A.~Shirazi, and K.~van Laerhoven, ``Are you in bed with
  technology?'' \emph{Pervasive Computing, IEEE}, vol.~11, no.~4, pp. 4--7,
  2012.

\bibitem{ContextAwareness_Holmquist_2001}
L.~E. Holmquist, F.~Mattern, B.~Schiele, , P.~Alahuhta, M.~Beigl, and H.~W.
  Gellersen, ``Smart-its friends: A technique for users to easily establish
  connections between smart artefacts,'' in \emph{Proceedings of the 3rd
  International Conference on Ubiquitous Computing}, 2001.

\bibitem{CommunicationTechnology_Friis_2946}
H.~Friis, \emph{Proceedings of the IRE}, vol.~34, p. 254, 1946.

\bibitem{Pervasive_Jiang_2012}
Y.~Jiang, X.~Pan, K.~Li, Y.~Lv, R.~P. Dick, M.~Hannigan, and L.~Shang, ``Ariel:
  Automatic wi-fi based room fingerprinting for indoor localization,'' in
  \emph{Proceedings of the 14th ACM International Conference on Ubiquitous
  Computing (Ubicomp 2012)}, 2012.

\bibitem{Pervasive_Pulkkinen_2012}
T.~Pulkkinen and P.~Nurmi, ``Awesom: Automatic discrete partitioning of indoor
  spaces for wifi fingerprinting,'' in \emph{Proceedings of the 10th
  International Conference on Pervasive Computing}, 2012.

\bibitem{DeviceFreeRecognition_Aly_2013}
H.~Aly and M.~Youssef, ``New insights into wifi-based device-free
  localization,'' in \emph{Adjunct Proceedings of the 2013 ACM International
  Joint Conference on Pervasive and Ubiquitous Computing (UbiComp 2013)}, ser.
  UbiComp '13, 2013.

\bibitem{Pervasive_Schougaard_2012}
K.~R. Schougaard, K.~Gronbaek, and T.~Scharling, ``Indoor pedestrian navigation
  based on hybrid route planning and location modelling,'' in \emph{Proceedings
  of the 10th International Conference on Pervasive Computing (Pervasive2012)},
  2012.

\bibitem{Pervasive_Wang_2012}
H.~Wang, S.~Sen, A.~Elgohary, M.~Farid, M.~Youssef, and R.~R. Choudhury, ``No
  need to war-drive -- unsupervised indoor localization,'' in \emph{Proceedings
  of the 10th International Conference on Mobile Systems, Applications and
  Services (Mobisys2012)}, 2012.

\bibitem{Pervasive_Chen_2012b}
Y.~Chen, D.~Lymberopoulos, J.~Liu, and B.~Priyantha, ``Fm-based indoor
  localisation,'' in \emph{Proceedings of the 10th International Conference on
  Mobile Systems, Applications and Services (Mobisys2012)}, 2012.

\bibitem{RFSensing_Sen_2012}
S.~Sen, B.~Radunovic, R.~R. Choudhury, and T.~Minka, ``You are facing the mona
  lisa -- spot localization using phy layer information,'' in \emph{Proceedings
  of the 10th International Conference on Mobile Systems, Applications and
  Services (Mobisys2012)}, 2012.

\bibitem{Pervasive_Scholz_2011b}
M.~Scholz, S.~Sigg, H.~R. Schmidtke, and M.~Beigl, ``Challenges for device-free
  radio-based activity recognition,'' in \emph{Proceedings of the 3rd workshop
  on Context Systems, Design, Evaluation and Optimisation, in conjunction with
  MobiQuitous 2011}, 2011.

\bibitem{Pervasive_Kosba_2011}
A.~E. Kosba, A.~Saeed, and M.~Youssef, ``Rasid: A robust wlan device-free
  passive motion detection system,'' in \emph{IEEE International Conference on
  Pervasive Computing and Communications (PerCom)}, 2012.

\bibitem{RFSensing_Lee_2010}
P.~W.~Q. Lee, W.~K.~G. Seah, H.-P. Tan, and Z.~Yao, ``Wireless sensing without
  sensors - an experimental study of motion/intrusion detection using rf
  irregularity,'' \emph{Measurement science and technology}, vol.~21, 2010.

\bibitem{DeviceFreeRecognition_Popleteev_2013}
A.~Popleteev, ``Device-free indoor localization using ambient radio systems,''
  in \emph{Adjunct Proceedings of the 2013 ACM International Joint Conference
  on Pervasive and Ubiquitous Computing (UbiComp 2013)}, ser. UbiComp '13,
  2013.

\bibitem{DeviceFreeRecognition_Lieckfeldt_2009}
D.~Lieckfeldt, J.~You, and D.~Timmermann, ``Characterizing the influence of
  human presence on bistatic passive rfid-system,'' in \emph{IEEE International
  Conference on Wireless and Mobile Computing, Networking and Communications},
  2009.

\bibitem{RFSensing_Mostofi_2013b}
Y.~Mostofi, ``Cooperative wireless-based obstacle/object mapping and
  see-through capabilities in robotic networks,'' \emph{Mobile Computing, IEEE
  Transactions on}, vol.~12, no.~5, pp. 817--829, 2013.

\bibitem{RFSensing_Mostofi_2011}
------, ``Compressive cooperative sensing and mapping in mobile networks,''
  \emph{Mobile Computing, IEEE Transactions on}, vol.~10, no.~12, pp.
  1769--1784, 2011.

\bibitem{RFSensing_Wilson_2009}
J.~Wilson and N.~Patwari, ``See-through walls: Motion tracking using
  variance-based radio tomography,'' \emph{IEEE Transactions on Mobile
  Computing}, vol.~10, no.~5, pp. 612--621, 2011.

\bibitem{RFSensing_Wilson_2010}
------, ``Radio tomographic imaging with wireless networks,'' \emph{IEEE
  Transactions on Mobile Computing}, vol.~9, pp. 621--632, 2010.

\bibitem{RFSensing_Bocca_2013}
M.~Bocca, A.~Luong, N.~Patwari, and T.~Schmid, ``Dial it in: Rotating rf
  sensors to enhance radio tomography,'' \emph{arXiv preprint arXiv:1312.5480},
  2013.

\bibitem{4022}
S.~Sigg, R.~M.~E. Masri, and M.~Beigl, ``Feedback based closed-loop carrier
  synchronisation: A sharp asymptotic bound, an asymptotically optimal
  approach, simulations and experiments,'' \emph{Transactions on mobile
  computing}, vol.~10, no.~11, pp. 1605--1617, November 2011.

\bibitem{DistributedBeamforming_Quitin_2013}
F.~Quitin, M.~M.~U. Rahman, R.~Mudumbai, and U.~Madhow, ``A scalable
  architecture for distributed transmit beamforming with commodity radios:
  Design and proof of concept,'' \emph{IEEE Transactions on Wireless
  Communications}, vol.~PP, no.~99, pp. 1--11, accepted for inclusion in a
  future issue.

\bibitem{DeviceFreeRecognition_Wagner_2012}
B.~Wagner and N.~Patwari, ``Passive rfid tomographic imaging for device-free
  user localization,'' in \emph{Workshop of Positioning,
  NavigationCommunication}, 2012.

\bibitem{DeviceFreeRecognition_Wagner_2013-2}
B.~Wagner, B.~Striebing, and D.~Timmermann, ``A system for live localization in
  smart environments,'' in \emph{IEEE International Conference on Networking,
  Sensing and Control}, 2013.

\bibitem{DeviceFreeRecognition_Wagner_2013}
B.~Wagner and D.~Timmermann, ``Adaptive clustering for device-free user
  positioning utilizing passive rfid,'' in \emph{Adjunct Proceedings of the
  2013 ACM International Joint Conference on Pervasive and Ubiquitous Computing
  (UbiComp 2013)}, ser. UbiComp '13, 2013.

\bibitem{CompressiveSensing_Candes_2006}
E.~Candes, J.~Romberg, and T.~Tao, ``Robust uncertainty principles: exact
  signal reconstruction from highly incomplete frequency information,''
  \emph{Information Theory, IEEE Transactions on}, vol.~52, no.~2, pp.
  489--509, 2006.

\bibitem{CompressiveSensing_Donoho_2006}
D.~Donoho, ``Compressed sensing,'' \emph{Information Theory, IEEE Transactions
  on}, vol.~52, no.~4, pp. 1289--1306, 2006.

\bibitem{CompressiveSensing_Needell_2010}
D.~Needell and R.~Vershynin, ``Signal recovery from incomplete and inaccurate
  measurements via regularized orthogonal matching pursuit,'' \emph{Selected
  Topics in Signal Processing, IEEE Journal of}, vol.~4, no.~2, pp. 310--316,
  2010.

\bibitem{RFSensing_Gonzalez_2013}
A.~Gonzalez-Ruiz and Y.~Mostofi, ``Cooperative robotic structure mapping using
  wireless measurements; a comparison of random and coordinated sampling
  patterns,'' \emph{Sensors Journal, IEEE}, vol.~13, no.~7, pp. 2571--2580,
  2013.

\bibitem{RFSensing_Gonzalez_2014}
A.~Gonzalez-Ruiz, A.~Ghaffarkhah, and Y.~Mostofi, ``An integrated framework for
  obstacle mapping with see-through capabilities using laser and wireless
  channel measurements,'' \emph{Sensors Journal, IEEE}, vol.~14, no.~1, pp.
  25--38, 2014.

\bibitem{RFSensing_Mostofi_2013}
Y.~Mostofi, ``Cooperative wireless-based obstacle/object mapping and
  see-through capabilities in robotic networks,'' \emph{Mobile Computing, IEEE
  Transactions on}, vol.~12, no.~5, pp. 817--829, 2013.

\bibitem{DeviceFreeRecognition_Wagner_2012-2}
B.~Wagner, D.~Timmermann, G.~Ruscher, and T.~Kirste, ``Device-free user
  localization utilizing neural networks and passive rfid,'' in
  \emph{Conference on Ubiquitous Positioning Indoor Navigation and Location
  Based Service (UPINLBS)}, 2012.

\bibitem{Pervasive_Jeon_2013}
S.~Jeon, Y.-J. Suh, C.~Yu, and D.~Han, ``Fast and accurate wi-fi localization
  in large-scale indoor venues,'' in \emph{Proceedings of the 10th
  International Conference on Mobile and Ubiquitous Systems: Computing,
  Networking and Services}, 2013.

\bibitem{Pervasive_Seifeldin_2009}
M.~Seifeldin and M.~Youssef, ``Nuzzer: A large-scale device-free passive
  localization system for wireless environments,'' \emph{CoRR}, vol.
  abs/0908.0893, 2009.

\bibitem{Pervasive_Seifeldin_2013}
M.~Seifeldin, A.~Saeed, A.~Kosba, A.~El-Keyi, and M.~Youssef, ``Nuzzer: A
  large-scale device-free passive localization system for wireless
  environments,'' \emph{IEEE Transactions on Mobile Computing (TMC)}, vol.~12,
  no.~7, 2013.

\bibitem{DeviceFreeRecognition_Zhang_2007}
D.~Zhang, J.~Ma, Q.~Chen, and L.~M. Ni, ``An rf-based system for tracking
  transceiver-free objects,'' in \emph{Proceedings of the 5th Annual IEEE
  International Conference on Pervasive Computing and Communications (PerCom
  2007)}, 2007.

\bibitem{RFSensing_Zhang_2009}
D.~Zhang and L.~Ni, ``Dynamic clustering for tracking multiple transceiver-free
  objects,'' in \emph{Proceedings of the 7th IEEE International Conference on
  Pervasive Computing and Communications}, 2009.

\bibitem{RFSensing_Lieckfeldt_2009}
D.~Lieckfeldt, J.~You, and D.~Timmermann, ``{Passive Tracking of
  Transceiver-Free Users with RFID},'' in \emph{{Intelligent Interactive
  Assistance and Mobile Multimedia Computing}}.\hskip 1em plus 0.5em minus
  0.4em\relax Springer Berlin Heidelberg, 2009, vol.~53, pp. 319--329.

\bibitem{RFSensing_Lieckfeldt_2009b}
------, ``Exploiting rf-scatter: Human localization with bistatic passive uhf
  rfid-systems,'' in \emph{Wireless and Mobile Computing, Networking and
  Communications, 2009. WIMOB 2009. IEEE International Conference on}, 2009,
  pp. 179--184.

\bibitem{RFSensing_Patwari_2011}
N.~Patwari and J.~Wilson, ``Spatial models for human motion-induced signal
  strength variance on static links,'' \emph{IEEE Transactions on Information
  Forensics and Security}, vol.~6, no.~3, pp. 791--802, September 2011.

\bibitem{Pervasive_Zhang_2012}
D.~Zhang, Y.~Liu, X.~Guo, M.~Gao, and L.~M. Ni, ``On distinguishing the
  multiple radio paths in rss-based ranging,'' in \emph{Proceedings of the 31st
  IEEE International Conference on Computer Communications}, 2012.

\bibitem{Pervasive_Scholz_2011}
M.~Scholz, S.~Sigg, D.~Shihskova, G.~von Zengen, G.~Bagshik, T.~Guenther,
  M.~Beigl, and Y.~Ji, ``Sensewaves: Radiowaves for context recognition,'' in
  \emph{Video Proceedings of the 9th International Conference on Pervasive
  Computing (Pervasive 2011)}, Jun. 2011.

\bibitem{4036}
M.~Reschke, J.~Starosta, S.~Schwarzl, and S.~Sigg, ``Situation awareness based
  on channel measurements,'' in \emph{Vehicular Technology Conference (VTC
  Spring), 2011 IEEE 73rd}, 2011.

\bibitem{ContextAwareness_Sigg_2011}
M.~Reschke, S.~Schwarzl, J.~Starosta, S.~Sigg, and M.~Beigl, ``Context
  awareness through the rf-channel,'' in \emph{Proceedings of the 2nd workshop
  on Context-Systems Design, Evaluation and Optimisation}, 2011.

\bibitem{OrganicComputing_Sigg_2011}
S.~Sigg, M.~Beigl, and B.~Banitalebi, \emph{Efficient adaptive communication
  from multiple resource restricted transmitters}, ser. Organic Computing - A
  Paradigm Shift for Complex Systems, Autonomic Systems Series.\hskip 1em plus
  0.5em minus 0.4em\relax Springer, 2011, ch. 5.4.

\bibitem{RFsensing_Xu_2013}
C.~Xu, B.~Firner, R.~S. Moore, Y.~Zhang, W.~Trappe, R.~Howard, F.~Zhang, and
  N.~An, ``Scpl: Indoor device-free multi-subject counting and localization
  using radio signal strength,'' in \emph{The 12th ACM/IEEE Conference on
  Information Processing in Sensor Networks (ACM/IEEE IPSN)}, 2013.

\bibitem{Pervasive_Sigg_2012}
S.~Sigg, M.~Scholz, S.~Shi, Y.~Ji, and M.~Beigl, ``Rf-sensing of activities
  from non-cooperative subjects in device-free recognition systems using
  ambient and local signals,'' \emph{IEEE Transactions on Mobile Computing},
  vol.~13, no.~4, 2013.

\bibitem{Pervasive_Sigg_2013}
S.~Sigg, S.~Shi, F.~Buesching, Y.~Ji, and L.~Wolf, ``Leveraging rf-channel
  fluctuation for activity recognition,'' in \emph{Proceedings of the 11th
  International Conference on Advances in Mobile Computing and Multimedia
  (MoMM2013)}, 2013.

\bibitem{DeviceFreeRecognition_Sigg_2013}
S.~Sigg, S.~Shi, and Y.~Ji, ``Rf-based device-free recognition of
  simultaneously conducted activities,'' in \emph{Adjunct Proceedings of the
  2013 ACM International Joint Conference on Pervasive and Ubiquitous Computing
  (UbiComp 2013)}, ser. UbiComp '13, 2013.

\bibitem{Pervasive_Sigg_2014b}
------, ``Teach your wifi-device: Recognise gestures and simultaneous
  activities from time-domain rf-features,'' \emph{International Journal on
  Ambient Computing and Intelligence (IJACI)}, vol.~6, no.~1, 2014.

\bibitem{Pervasive_Shi_2012}
S.~Shi, S.~Sigg, and Y.~Ji, ``Passive detection of situations from ambient
  fm-radio signals,'' in \emph{Proceedings of the 2012 ACM Conference on
  Ubiquitous Computing}, ser. UbiComp '12, 2012.

\bibitem{Pervasive_Shi_2012b}
------, ``Activity recognition from radio frequency data: Multi-stage
  recognition and features,'' in \emph{IEEE Vehicular Technology Conference
  (VTC Fall)}, 2012.

\bibitem{DeviceFreeRecognition_Shi_2013}
------, ``Joint localisation and activity recognition from ambient fm broadcast
  signals,'' in \emph{Adjunct Proceedings of the 2013 ACM International Joint
  Conference on Pervasive and Ubiquitous Computing (UbiComp 2013)}, ser.
  UbiComp '13, 2013.

\bibitem{Pervasive_Scholz_2013}
M.~Scholz, T.~Riedel, M.~Hock, and M.~Beigl, ``Device-free and device-bound
  activity recognition using radio signal strength,'' in \emph{Proceedings of
  the 4th Augmented Human International Conference in cooperation with ACM
  SIGCHI}, 2013.

\bibitem{Pervasive_Sigg_2014}
S.~Sigg, M.~Hock, M.~Scholz, G.~Troester, L.~Wolf, Y.~Ji, and M.~Beigl,
  ``Passive, device-free recognition on your mobile phone: tools, features and
  a case study,'' in \emph{Proceedings of the 10th International Conference on
  Mobile and Ubiquitous Systems: Computing, Networking and Services}, 2013.

\bibitem{RFsensing_Kim_2009}
Y.~Kim and H.~Ling, ``Human activity classification based on micro-doppler
  signatures using a support vector machine,'' \emph{IEEE Transactions on
  Geoscience and Remote Sensing}, vol.~47, no.~5, pp. 1328--1337, 2009.

\bibitem{RFSensing_Kellog_2014}
\BIBentryALTinterwordspacing
``Bringing gesture recognition to all devices,'' in \emph{Presented as part of
  the 11th USENIX Symposium on Networked Systems Design and
  Implementation}.\hskip 1em plus 0.5em minus 0.4em\relax Berkeley, CA: USENIX,
  2014. [Online]. Available:
  \url{https://www.usenix.org/conference/nsdi14/technical-sessions/
  presentation/kellogg}
\BIBentrySTDinterwordspacing

\bibitem{RFSensing_Adib_2014}
F.~Adib, Z.~Kabelac, D.~Katabi, and R.~C. Miller, ``3d tracking via body radio
  reflections,'' available online:
  http://witrack.csail.mit.edu/witrack-paper.pdf, 2014, usenix NSDI’14 - to
  appear.

\bibitem{icdar1}
K.~Kunze, H.~Kawaichi, K.~Yoshimura, and K.~Kise, ``The wordmeter -- estimating
  the number of words read using document image retrieval and mobile eye
  tracking,'' in \emph{Proc. ICDAR 2013}, 2013.

\bibitem{cunningham2001reading}
A.~Cunningham and K.~Stanovich, ``What reading does for the mind,''
  \emph{Journal of Direct Instruction}, vol.~1, no.~2, pp. 137--149, 2001.

\bibitem{kunze2013know}
K.~Kunze, Y.~Utsumi, Y.~Shiga, K.~Kise, and A.~Bulling, ``I know what you are
  reading: recognition of document types using mobile eye tracking,'' in
  \emph{Proceedings of the 17th annual international symposium on International
  symposium on wearable computers}.\hskip 1em plus 0.5em minus 0.4em\relax ACM,
  2013, pp. 113--116.

\bibitem{Buscher:2008:GUG:1358628.1358805}
\BIBentryALTinterwordspacing
G.~Buscher, A.~Dengel, L.~van Elst, and F.~Mittag, ``Generating and using
  gaze-based document annotations,'' in \emph{CHI '08 Extended Abstracts on
  Human Factors in Computing Systems}, ser. CHI EA '08.\hskip 1em plus 0.5em
  minus 0.4em\relax New York, NY, USA: ACM, 2008, pp. 3045--3050. [Online].
  Available: \url{http://doi.acm.org/10.1145/1358628.1358805}
\BIBentrySTDinterwordspacing

\bibitem{kunze2013wordometer}
K.~Kunze, H.~Kawaichi, K.~Yoshimura, and K.~Kise, ``The wordometer--estimating
  the number of words read using document image retrieval and mobile eye
  tracking,'' in \emph{Document Analysis and Recognition (ICDAR), 2013 12th
  International Conference on}.\hskip 1em plus 0.5em minus 0.4em\relax IEEE,
  2013, pp. 25--29.

\bibitem{kunze2014}
A.~Okoso, K.~Kunze, and K.~Kise, ``Implicit gaze based annotations to support
  second language learning,'' in \emph{Proceedings of the 2014 ACM Conference
  on Pervasive and Ubiquitous Computing Adjunct Publication}, ser. UbiComp '13
  Adjunct, 2014, pp. 155--158.

\bibitem{ishimaru2014blink}
S.~Ishimaru, K.~Kunze, K.~Kise, J.~Weppner, A.~Dengel, P.~Lukowicz, and
  A.~Bulling, ``In the blink of an eye - combining head motion and eye blink
  frequency for activity recognition with google glass,'' in \emph{Proceedings
  of the 5th Augmented Human International Conference}.\hskip 1em plus 0.5em
  minus 0.4em\relax ACM, 2014, pp. 150--153.

\bibitem{Pervasive_Boano_2014}
C.~A. Boano, M.~Z{\'u}{\~n}iga, J.~Brown, U.~Roedig, C.~Keppitiyagama, and
  K.~R{\"o}mer, ``Templab: a testbed infrastructure to study the impact of
  temperature on wireless sensor networks,'' in \emph{Proceedings of the 13th
  international symposium on Information processing in sensor networks}.\hskip
  1em plus 0.5em minus 0.4em\relax IEEE Press, 2014, pp. 95--106.

\bibitem{RFSensing_Ruddle_2003}
A.~R. Ruddle, D.~A. Topham, and D.~D. Ward, ``Investigation of electromagnetic
  emissions measurements practices for alternative powertrain road vehicles,''
  in \emph{Electromagnetic Compatibility, 2003 IEEE International Symposium
  on}, vol.~2.\hskip 1em plus 0.5em minus 0.4em\relax IEEE, 2003, pp. 543--547.

\bibitem{RFSensing_CISPR12_1997}
``Vehicles, motorboats and devices-radio, spark-ignited engine-driven devices.
  radio disturbance characteristics--limits and methods of measurement,'' 1997.

\bibitem{RFSensing_CISPR22_1999}
``Limits and measurement of electromagnetic disturbance characteristics of
  information technology equipment,'' 1999.

\bibitem{RFSensing_Dong_2006}
X.~Dong, H.~Weng, D.~G. Beetner, T.~H. Hubing, D.~C. Wunsch, M.~Noll, H.~Goksu,
  and B.~Moss, ``Detection and identification of vehicles based on their
  unintended electromagnetic emissions,'' \emph{Electromagnetic Compatibility,
  IEEE Transactions on}, vol.~48, no.~4, pp. 752--759, 2006.

\bibitem{RFSensing_Kassem_2012}
N.~Kassem, A.~Kosba, and M.~Youssef, ``Rf-based vehicle detection and speed
  estimation,'' in \emph{75th IEEE Vehicular Technology Conference (VTC
  Spring)}, 2012, pp. 1--5.

\bibitem{RFSensing_Ding_2012}
Y.~Ding, B.~Banitalebi, T.~Miyaki, and M.~Beigl, ``Rftraffic: a study of
  passive traffic awareness using emitted rf noise from the vehicles,''
  \emph{EURASIP Journal on Wireless Communications and Networking}, vol. 2012,
  no.~1, pp. 1--14, 2012.

\bibitem{Emotion_Nguyen_2012}
M.~R. Anh-Tuan~Nguyen, Wei~Chen, ``The role of human body expression in affect
  detection: A review,'' in \emph{10th Asia Pacific Conference on Computer
  Human Interaction (APCHI 2012)}, 2012.

\bibitem{Emotion_Castellano_2008}
G.~Castellano, L.~Kessous, and G.~Caridakis, ``Emotion recognition through
  multiple modalities: face, body gesture, speech,'' in \emph{Affect and
  emotion in human-computer interaction}.\hskip 1em plus 0.5em minus
  0.4em\relax Springer, 2008, pp. 92--103.

\bibitem{Emotion_Meeren_2005}
H.~K. Meeren, C.~C. van Heijnsbergen, and B.~de~Gelder, ``Rapid perceptual
  integration of facial expression and emotional body language,''
  \emph{Proceedings of the National Academy of Sciences of the United States of
  America}, vol. 102, no.~45, pp. 16\,518--16\,523, 2005.

\bibitem{Emotion_Walters_1986}
K.~Walters and R.~Walk, ``Perception of emotion from body posture,''
  \emph{Bulletin of the Psychonomic Society}, vol.~24, no.~5, 1986.

\bibitem{Emotion_Dittmann_1978}
A.~T. Dittmann, ``The role of body movement in communication,'' \emph{Nonverbal
  behavior and communication}, pp. 69--95, 1978.

\bibitem{Emotion_Wallbott_1998}
H.~G. Wallbott, ``Bodily expression of emotion,'' \emph{European journal of
  social psychology}, vol.~28, no.~6, pp. 879--896, 1998.

\bibitem{Emotion_VanHeijnsbergen_2007}
C.~Van~Heijnsbergen, H.~Meeren, J.~Grezes, and B.~de~Gelder, ``Rapid detection
  of fear in body expressions, an erp study,'' \emph{Brain research}, vol.
  1186, pp. 233--241, 2007.

\bibitem{Emotion_Atkinson_2007}
A.~P. Atkinson, M.~L. Tunstall, and W.~H. Dittrich, ``Evidence for distinct
  contributions of form and motion information to the recognition of emotions
  from body gestures,'' \emph{Cognition}, vol. 104, no.~1, pp. 59--72, 2007.

\bibitem{Emotion_Bull_1987}
P.~E. Bull, \emph{Posture and gesture.}\hskip 1em plus 0.5em minus 0.4em\relax
  Pergamon press, 1987.

\bibitem{Emotion_DeMeijer_1989}
M.~De~Meijer, ``The contribution of general features of body movement to the
  attribution of emotions,'' \emph{Journal of Nonverbal behavior}, vol.~13,
  no.~4, pp. 247--268, 1989.

\bibitem{Emotion_Montepare_1999}
J.~Montepare, E.~Koff, D.~Zaitchik, and M.~Albert, ``The use of body movements
  and gestures as cues to emotions in younger and older adults,'' \emph{Journal
  of Nonverbal Behavior}, vol.~23, no.~2, pp. 133--152, 1999.

\bibitem{Emotion_Crane_2007}
E.~Crane and M.~Gross, ``Motion capture and emotion: Affect detection in whole
  body movement,'' in \emph{Affective computing and intelligent
  interaction}.\hskip 1em plus 0.5em minus 0.4em\relax Springer, 2007, pp.
  95--101.

\bibitem{Emotion_Bernhardt_2007}
D.~Bernhardt and P.~Robinson, ``Detecting affect from non-stylised body
  motions,'' in \emph{Affective Computing and Intelligent Interaction}.\hskip
  1em plus 0.5em minus 0.4em\relax Springer, 2007, pp. 59--70.

\bibitem{Emotion_Lagerlof_2009}
I.~Lagerl{\"o}f and M.~Djerf, ``Children's understanding of emotion in dance,''
  \emph{European Journal of Developmental Psychology}, vol.~6, no.~4, pp.
  409--431, 2009.

\bibitem{AttentionMonitoring_Xu_2012}
Y.~Xu, N.~Stojanovic, L.~Stojanovic, and T.~Schuchert, ``Efficient human
  attention detection based on intelligent complex event processing,'' in
  \emph{Proceedings of the 6th ACM International Conference on Distributed
  Event-Based Systems}, ser. DEBS '12, 2012, pp. 379--380.

\bibitem{AttentionMonitoring_Wu_2007}
F.~Wu and B.~Hubermann, ``Novelty and collective attention,'' in
  \emph{Proceedings of the National Academics of Sciences}, vol. 104, no.~45,
  2007, pp. 17\,599--17\,601.

\bibitem{AttentionMonitoring_Wickens_1984}
C.~Wickens, \emph{Processing resources in attention}.\hskip 1em plus 0.5em
  minus 0.4em\relax Academic Press, 1984.

\bibitem{AttentionMonitoring_Yonezawa_2007}
T.~Yonezawa, H.~Yamazoe, A.~Utsumi, and S.~Abe, ``Gaze-communicative behavior
  of stuffed-toy robot with joint attention and eye contact based on ambient
  gaze-tracking,'' in \emph{ICMI}, 2007, pp. 140--145.

\bibitem{AttentionMonitoring_Wickens_2008}
C.~Wickens and J.~McCarley, \emph{Applied attention theory}.\hskip 1em plus
  0.5em minus 0.4em\relax CRC Press, 2008.

\bibitem{AttentionMonitoring_Gollan_2011}
B.~Gollan, B.~Wally, and A.~Ferscha, ``Automatic attention estimation in an
  interactive system based on behaviour analysis,'' in \emph{Proceedings of the
  15th Portuguese Conference on Artificial Intelligence (EPIA2011)}, 2011.

\bibitem{AttentionMonitoring_Ferscha_2012}
A.~Ferscha, K.~Zia, and B.~Gollan, ``Collective attention through public
  displays,'' in \emph{2012 IEEE Sixth International Conference on
  Self-Adaptive and Self-Organizing Systems (SASO)}, 2012, pp. 211 --216.

\bibitem{RFSensing_Wu_2012}
K.~Wu, J.~Xiao, Y.~Yi, M.~Gao, and L.~M. Ni, ``Fila: Fine-grained indoor
  localization,'' in \emph{INFOCOM, 2012 Proceedings IEEE}.\hskip 1em plus
  0.5em minus 0.4em\relax IEEE, 2012, pp. 2210--2218.

\bibitem{RFSensing_yang_2013}
Z.~Yang, Z.~Zhou, and Y.~Liu, ``From rssi to csi: Indoor localization via
  channel response,'' \emph{ACM Computing Surveys (CSUR)}, vol.~46, no.~2,
  p.~25, 2013.

\bibitem{RFSensing_Nerguizian_2006}
C.~Nerguizian, C.~Despins, and S.~Aff{\`e}s, ``Geolocation in mines with an
  impulse response fingerprinting technique and neural networks,''
  \emph{Wireless Communications, IEEE Transactions on}, vol.~5, no.~3, pp.
  603--611, 2006.

\bibitem{RFSensing_Patwari_2007}
N.~Patwari and S.~K. Kasera, ``Robust location distinction using temporal link
  signatures,'' in \emph{Proceedings of the 13th annual ACM international
  conference on Mobile computing and networking}.\hskip 1em plus 0.5em minus
  0.4em\relax ACM, 2007, pp. 111--122.

\bibitem{RFSensing_Zhang_2008}
J.~Zhang, M.~H. Firooz, N.~Patwari, and S.~K. Kasera, ``Advancing wireless link
  signatures for location distinction,'' in \emph{Proceedings of the 14th ACM
  international conference on Mobile computing and networking}.\hskip 1em plus
  0.5em minus 0.4em\relax ACM, 2008, pp. 26--37.

\bibitem{RFSensing_Halperin_2011}
D.~Halperin, W.~Hu, A.~Sheth, and D.~Wetherall, ``Predictable 802.11 packet
  delivery from wireless channel measurements,'' \emph{ACM SIGCOMM Computer
  Communication Review}, vol.~41, no.~4, pp. 159--170, 2011.
\end{thebibliography}

%

\begin{IEEEbiography}[{\includegraphics[width=1in,height=1.25in,clip,keepaspectratio]{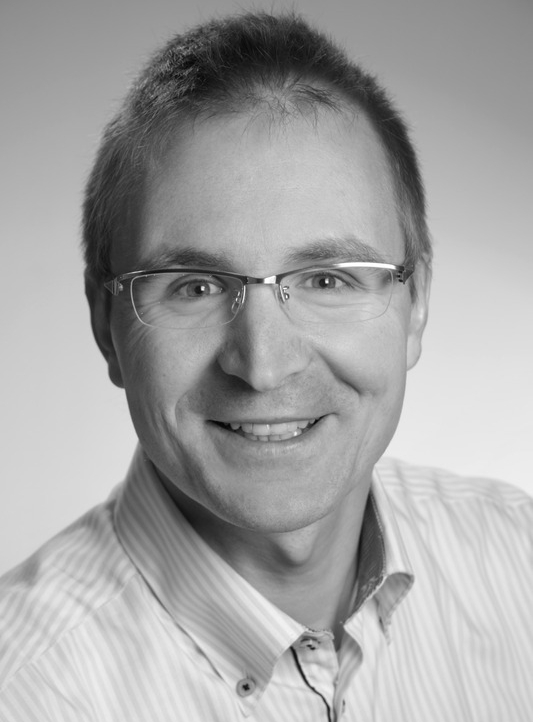}}]{Stephan Sigg}
is with the Georg-August University Goettingen.  
Before, he was with Karlsruhe Institute of Technology (2010) and TU Braunschweig (2007-2010).
He as been an academic guest at Helsinki University (2014) and at ETH Zurich (2013) as well as a guest researcher at National Institute of Informatics, Japan (2010-2013).
He obtained his diploma in computer science from Univ. of Dortmund in 2004 and finished his PhD in 2008 at the chair for communication technology at the Univ. of Kassel.
His research interests include the design, analysis and optimisation of algorithms for context aware and Ubiquitous systems.
\end{IEEEbiography}

\begin{IEEEbiography}[{\includegraphics[width=1in,height=1.25in,clip,keepaspectratio]{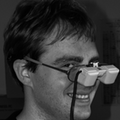}}]{Kai Kunze}
works as an assistant professor at the Intelligent Media Processing Group, Osaka Prefecture University, directed by Prof. Koichi Kise and senior researcher affiliated with Keio Media Design. 
He received a Summa Cum Laude for his phD thesis, University Passau. 
He was visting researcher at the MIT Media Lab. 
His work experience includes internships at the Palo Alto Research Center (PARC), Sunlabs Europe and the Research Department of the German Stock Exchange. 
His major research contributions are in pervasive computing, especially in sensing, phyisical and cognitive activity recognition. 
Recently, he focuses on tracking knowledge acquisition activities, especially reading.
\end{IEEEbiography}

\begin{IEEEbiography}[{\includegraphics[width=1in,height=1.25in,clip,keepaspectratio]{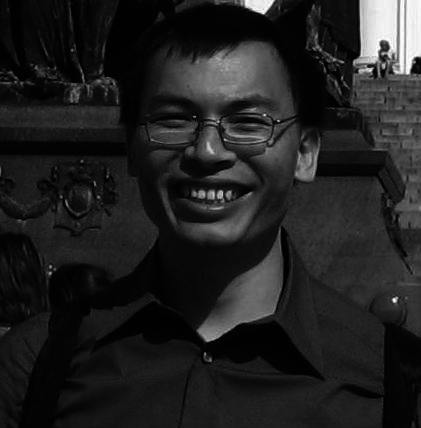}}]{Xiaoming Fu}
received his Ph.D. in computer science from Tsinghua University, China 
in 2000.
He was a research staff at Technical University Berlin until joining the 
University of G\"{o}ttingen, Germany in 2002, where he has been a full 
professor and heads the computer networks group since 2007. His research 
interests are architectures, protocols, and applications for networked 
systems, including information dissemination, mobile systems, cloud 
computing and social networks.
\end{IEEEbiography}




\end{document}